\documentclass{aa}
\usepackage{graphicx}
\usepackage{txfonts}
\usepackage{supertabular}
\usepackage{longtable}
\usepackage{rotating}
\usepackage{lscape}
\begin{document}

   \title{A Method to Derive the Absolute Composition of the Sun, the Solar System 
          and the Stars}

   \author{Luciano Piersanti
           \inst{1},
           Oscar Straniero
           \inst{1}
           \&
           Sergio Cristallo
           \inst{1}
          }

   \offprints{L. Piersanti}

   \institute{INAF-Osservatorio Astronomico di Collurania Teramo
              via Mentore Maggini, snc, 64100, Teramo-IT\\
              \email{cristallo,piersanti,straniero@oa-teramo.inaf.it}
             }

   \date{}

  \abstract
    {
     The knowledge of isotopic and elemental abundances of the pristine solar system material 
     provides a fundamental test of galactic chemical evolution models, while the composition 
     of the solar photosphere is a reference pattern to understand stellar abundances. However, 
     spectroscopic or meteoritic abundance determinations are only possible for an incomplete 
     sample of the 83 elements detected in the solar system. Therefore, only relative 
     abundances are experimentally determined, with respect to $\element[][]{H}$ or to 
     $\element[][]{Si}$ for spectroscopic or meteoritic measurements, respectively. 
     For this reason, the available compilations of solar abundances are obtained by combining 
     spectroscopic and meteoritic determinations, a procedure requiring the knowledge of the 
     chemical modification occurred in the solar photosphere.
    } 
    {
     We provide a method to derive the mass fractions of all the 83 elements (and their most 
     abundant isotopes) in the early solar system material and in the present-day solar surface. 
    }
    {
     By computing a solar model, we investigate physical processes responsible for the variation 
     of the solar surface composition in last 4.75 Gyr. An extended network, from 
     $\element[][]{H}$ to $\element[][]{U}$, is coupled to our stellar evolutionary code. The 
     effects of microscopic diffusion, rotational-induced mixing in the tachocline and 
     radiative acceleration are discussed. 
    }
    {
     The abundances of all the 83 elements are given for both the pristine solar system and 
     the solar photosphere. Calculations are repeated by adopting the most widely adopted 
     compilations of solar abundances. Since for a given $\mathrm{[Fe/H]}$, the total 
     metallicity depends on $(Z/X)_\odot$, a 30\% reduction of $Z$ is found when passing from 
     the classical Anders \& Grevesse to the most recent Lodders compilation. Some 
     implications are discussed, as, in particular, an increase of about 700 Myr of the 
     estimated age of Globular Clusters. 
    }
    {
     Within the experimental errors, the complete set of relative solar abundances, as 
     obtained by combining meteoritic and photospheric measurements, are consistent with the 
     variations implied by the quoted physical processes. Few deviations can be easily 
     attributed to the decay of long-lived radioactive isotopes. The huge lithium depletion 
     is only partially explained by introducing a rotational-induced mixing in the tachocline.
    }

   \titlerunning{Chemical abundances of stars}
   \authorrunning{L. Piersanti et al.}
   \maketitle

\section{Introduction}

The Sun is the only star for which a full set of elemental abundances is available. For this 
reason, its composition represents the reference framework for the abundance analyses in 
Astrophysics. In the standard spectroscopic notation, abundances are given with respect to 
the solar composition ({\it e.g.} 
$\mathrm{[Fe/H]}=\log\left(X_\mathrm{Fe}/X_\mathrm{H}\right)-\log\left(X_\mathrm{Fe}/X_\mathrm{H})_\odot\right)$.
Then, according to a widely accepted procedure, the abundances of the various elements are 
obtained by {\it re-scaling} the solar composition: a trace element (usually iron) is used 
to determine the scaling factor, which is applied to all the other elements, except 
$\element[][]{H}$ and $\element[][]{He}$. Thus, the distribution of heavy elements in the 
Sun is assumed as a sort of {\it cosmic} composition. Such a scenario obviously contrasts 
with many theoretical and observational evidences. 
We know that the distribution of heavy elements in the interstellar medium changes with time, 
as a consequence of the pollution caused by different generation of stars. 
The case of the enhancement of the $\alpha$ elements ($\element[][]{O}$, $\element[][]{Ne}$, 
$\element[][]{Mg}$, $\element[][]{Si}$, $\element[][]{S}$, $\element[][]{Ca}$, 
$\element[][]{Ti}$) in the galactic halo, where $\left[\alpha/\mathrm{Fe}\right]\sim 0.3-0.6$, 
represents a well known example of a deviation from the scaled-solar composition. A similar 
overabundance, also shown by the $r$-process elements, like Europium, is interpreted as an 
evidence of the fact that the early Galaxy, unlike the pre-solar nebulae, was mainly polluted 
by massive stars.
Notwithstanding these limitations, the solar composition remains an important test for models 
of galactic chemical evolution. Indeed, it represents a typical pattern for the evolved 
galactic composition, like the one shown by disk stars. 

Abundances of all the 83 elements found in nature have been obtained from various components 
of the solar system, like Earth, Moon, planets, comets, meteorites, solar wind, solar corona 
and, obviously, solar photosphere. For some elements, abundances in several solar system 
environments are available, whilst others can be measured just in one component. For example, 
heavy elements abundances are generally best measured in meteorites. On the contrary, 
$\element[][]{H}$ and other volatile elements (like $\element[][]{C}$, $\element[][]{N}$, 
$\element[][]{O}$) are strongly depleted in meteorites and their abundances are typically 
derived from solar spectra. In some cases, as for the noble gases $\element[][]{Ne}$ and 
$\element[][]{Ar}$, poor information can be found both in meteorites and in the solar 
photosphere, so that their abundances are derived from high energy spectra of the hot solar 
corona or from the solar wind. Finally, the abundances of $\element[][]{Kr}$ and 
$\element[][]{Xe}$ can not be directly measured and they are usually estimated by means of 
nucleosynthesis models.

Thus, the data set obtained from a single solar-system component is incomplete, {\it i.e.} 
it only includes a subsample of the 83 elements existing in the solar system. For this 
reason, only relative abundances\footnote{Hereinafter {\it relative abundances} indicates 
the abundance ratios, by mass or by number, while {\it absolute abundances} refers to mass 
or number fraction, $x_i$ or $n_i$, respectively.} can be obtained (see Section \ref{sec2}). 
Moreover, a great caution should be used when these relative abundances, as collected from 
different solar system environments, are combined to obtain a complete compilation.
As a matter of fact, changes with respect to the pristine composition have been produced by 
different physical processes (Anders \& Grevesse 1989; Grevesse \& Noels 1993; Grevesse 
\& Sauval 1998; Lodders 2003; Palme \& Jones 2004). 
For example, melting or crystallization could have produced a certain differentiation on 
planets and minor bodies. In this context, among the various meteoritic phases, CI 
carbonaceous chondrites represent the best primitive sample of the solar system composition. 
On the contrary, the composition of the solar photosphere has been substantially modified 
since the epoch of the solar system formation. Indeed, the solar surface is at the top of 
a deep convective region, which penetrates for about 30\% of the stellar radius. At the base 
of this region, where the temperature attains $\sim 2\cdot 10^6$ K, light isotopes ({\it e.g.} 
$\element[][]{D}$ or $\element[][]{Li}$) are burned via proton captures. It should be noted 
that the temperatures experienced within the convective layer of the Sun were even larger 
in the past 4.5 Gyr (up to $3\div 4\cdot 10^6$ K during the pre-Main Sequence). 
In addition, other phenomena, such as mass loss, eventually coupled with the external 
convection, may have contributed to the variation of the surface composition of the Sun. 
Indeed, the $\element[][]{He}$ abundance derived from the inversion of helioseismic data 
($Y_\odot=0.2485\pm 0.034$ - Basu \& Antia 2004) can be only explained by invoking 
gravitational settling or, more in general, microscopic diffusion. Note that diffusion 
affects the whole chemical pattern. In particular, the abundances of all the isotopes more 
massive than $A=12$ are depleted at the solar surface ($\sim$ -10\% with respect to the 
original composition), whilst the mass fraction of $\element[][]{H}$ increases (about +5\%).
For this reason, photospheric abundances, usually expressed relative to $\element[][]{H}$, 
cannot represent the original composition of the solar system. This occurrence should be 
carefully taken into account, especially for those elements whose abundances are almost 
exclusively measured in the solar photosphere ({\it e.g.} $\element[][]{C}$, $\element[][]{N}$, 
$\element[][]{O}$ and noble gases). Note that these elements represent about 50\% of the 
total metallicity ($Z$). 

In principle, a comparison between photospheric and meteoritic abundances could provide a 
direct evidence of the possible chemical modifications occurred at the solar surface during 
the last 4.5 Gyr. Such a comparison reveals a general agreement between photospheric and 
meteoritic (CI chondrites) relative abundances. In particular, according to Palme \& Jones 
(2004), the quantity $\left(n(\mathrm{el})/n(\mathrm{Si})\right)_\mathrm{PH}/
\left(n(\mathrm{el})/n(\mathrm{Si})\right)_\mathrm{CI}$ \footnote {``$n(\mathrm{el})$'' is 
the number abundance of a generic element, while ``$\mathrm{PH}$'' and ``$\mathrm{CI}$'' 
refers to photospheric and CI carbonaceous chondrites, respectively.} averaged for 34 
elements with spectroscopic uncertainty smaller than 25\% is $1.004 \pm 0.022$ 
\footnote{Here, the error of the mean has been reported.}.

The similarity of meteoritic and photospheric relative abundances does not necessarily imply 
that the absolute abundances have been maintained unchanged at the solar surface. Nonetheless, 
this occurrence represents an important constraint to the physical processes affecting the 
chemical composition of the outer layer of the Sun. For instance, these processes must 
preserve the abundance ratios, at least within the experimental errors. 
The most clear deviation from this rule is represented by lithium, whose abundance relative 
to $\element[][]{Si}$ is about 2 order of magnitude larger in meteorites than at the solar 
photosphere. Such a strong depletion of the photospheric $\element[][]{Li}$ is considered 
an evidence for nuclear burning (mainly via $p$-captures) occurred at the bottom of the 
solar convective envelope or immediately below it (D'Antona \& Mazzitelli 1984; Pinsonneault 
et al. 1989). 
Note that, since $p$-captures rates are significantly different from one isotope to another, 
they represent a typical example of physical process not preserving the relative abundances. 

In this paper, on the base of up to date Standard Solar Models (SSMs), we discuss the modifications 
occurred at the solar photosphere, since the pre-Main Sequence phase. We will show, in 
particular, that in SSMs where the modification of the surface composition is only caused by 
convective mixing and microscopic diffusion, relative abundances among heavy elements are 
preserved, with only few exceptions, and, therefore, they fulfill the constraints imposed by 
the comparison between meteoritic and photospheric relative abundances of the best measured 
refractory elements.

Note, however, that in many astrophysical applications and, in particular, in the calculation 
of stellar models, absolute, rather than relative, abundances are required, but there is no 
way to derive the solar absolute composition on the base of the available measurements only.  
Forgetting, for a while, the differences in the chemical composition of the various solar 
system components and adopting a compilation obtained by combining different data sources, 
one may obtain a partial solution of this problem. The mass fraction of $\element[][]{He}$ 
($Y$) of the present-day-solar photosphere can indeed be independently estimated according 
to helioseismic measurements. Then, the detailed absolute composition of the surface layers 
of the present Sun could be obtained by deriving the hydrogen mass fraction ($X$) from the 
relation $X+Y+Z=Y+(1+Z/X)\cdot X=1$, where $Z/X$ is known from the compilation of relative 
abundances. However, also in this case, the composition of the pristine solar system would 
remain undetermined. This problem can be solved by using a stellar evolution code to describe 
the past solar evolution. In this way, we are able to distinguish between present-day solar 
(photospheric) abundances and the composition of the pre-solar nebula. The photospheric 
abundances should be used to derive the composition of stars when, as usual, the results of 
abundance analyses are given relative to the Sun. At the same time, pristine abundances 
should be used to constrain the chemical evolution of the Galaxy.

Let us remark that our goal is the definition of a standard procedure for the derivation of 
the absolute composition of the Sun and the early Solar System. Since a critical analysis 
of the reliability of the available solar composition measurements is far from our expertise, 
we have repeated the derivation of the absolute abundances adopting the most used data sets. 
The main implications for various astrophysical problems requiring the knowledge of the solar 
composition will be discussed in Section \ref{sec5}. 
We do not address the largely debated question concerning the reliability of standard solar 
models. The reader can easily find in the recent literature a large number of papers on this 
issue ({\it e.g.} Bahcall, Serenelli \& Pinsonneault 2004; Bahcall et al. 2005, and references 
therein).

\section{Solar and Solar System Abundances}\label{sec2}

Solar (photospheric) abundances, as derived from spectroscopic analysis, are usually expressed 
in a logarithmic scale relative to hydrogen:
\begin{equation}
\varepsilon(\mathrm{el})=\log\left[{\frac{n(\mathrm{el})}{n(\mathrm{H})}}\right]+12
\label{eq:1}
\end{equation}
In this scale the hydrogen abundance is set, by definition, to $\varepsilon(\mathrm{H})=12$.

The elemental abundances of the pristine solar system material, mainly from CI carbonaceous 
chondrites, are usually given relative to silicon, whose abundance is fixed to $10^6$ atoms:
\begin{equation}
N(\mathrm{el})=\frac{n(\mathrm{el})}{n(\mathrm{Si})}\cdot 10^6
\label{eq:2}
\end{equation}

The solar and the solar system abundance scales are referred as {\it astronomical} and 
{\it cosmochemical} scales, respectively. As recalled in the Introduction, they are both 
incomplete, so that the available compilations of solar chemical abundances necessarily 
contain data from different sources. Following a commonly adopted procedure, the meteoritic 
abundances are ``translated'' from the cosmochemical into the astronomical scale by means of 
the following relation:
\begin{equation}
\varepsilon(\mathrm{el})=R+\log (N(\mathrm{el}))
\label{eq:3}
\end{equation}
Here the re-scaling factor $R$ is obtained by inverting Eq.~(\ref{eq:3}) for silicon, whose 
abundances in the cosmochemical and in the astronomical scales are firmly measured (Lodders 
2003), or by taking an average value over a selection of few elements detectable both in 
meteorites and in the solar photosphere ({\it e.g.} Cameron 1973, 1982; Grevesse 1984; Anders 
\& Grevesse 1989; Palme \& Beer 1993). Such a procedure was firstly proposed by Suess \& Urey 
(1956), according to the pioneer work of Clarke (1889), and it was based on the implicit 
assumption of the existence of a unique cosmochemical pattern. As recalled in the 
Introduction, we know that this statement is not correct.
Note, however, that the re-mapping procedure remains correct, provided that the evolution of 
the surface composition of the Sun preserves the relative abundance ratios. In this case, by 
summing and subtracting $\log\left(n(\mathrm{Si})\cdot 10^6\right)$ from the right side of 
Eq.~(\ref{eq:1}), one obtains:
\begin{equation}
\varepsilon(\mathrm{el})=\log{\left[\frac{n(\mathrm{el})}{n(\mathrm{Si})}\cdot 10^6\right]}+\log{\left[{\frac{n(\mathrm{Si})}{n(\mathrm{H})}}\cdot 10^{-6}\right]}+12
\label{eq:4}
\end{equation}
\noindent
Then, by using the definition of $N(\mathrm{el})$ in Eq.~(\ref{eq:2}) and putting
$R=\log{\left[{\frac{n(\mathrm{Si})}{n(\mathrm{H})}}\right]}+6$, we can recognize that the
last equation coincides with Eq.~(\ref{eq:3}). Therefore, if the evolution of the Sun 
preserves the ratios $n(\mathrm{el})/n(\mathrm{Si})$, the same (photospheric) abundance
$\varepsilon(\mathrm{el})$ can be obtained by using the definition in Eq.~(\ref{eq:1})
or in Eq.~(\ref{eq:3}).

In order to avoid an easy miss-interpretation, we have to keep in mind the previous
considerations in the practical use of the abundance compilations obtained by combining 
photospheric and meteoritic data. In particular, the most frequent misunderstanding concerns 
the comparison between the values of $\varepsilon(\mathrm{el})$ obtained from spectroscopy 
(astronomical scale, as defined by Eq.~(\ref{eq:1})) and those obtained by re-mapping the 
cosmochemical into the astronomical scale according to Eq.~(\ref{eq:3}). The similarity of 
the two scales is often interpreted as a problem for SSMs including microscopic diffusion, 
because they imply a depletion of the heavy elements at the solar photosphere (see {\it e.g.} 
Grevesse \& Sauval 1998). Such a conclusion has been questioned by Lodders (2003) by observing 
that microscopic diffusion could have affected by the same amount the abundances of heavy 
elements, so that the relative ratios would be preserved. In the following, we will 
demonstrate that this is the case.

If the isotopic, rather than the elemental, composition is needed, further difficulties 
should be considered. In some cases, precise isotopic ratios can be obtained for meteorites, 
but their spectroscopic determination is limited by the small isotopic shift of the atomic 
lines relative to their width. For this reason, the standard procedure makes large use of 
the very precise terrestrial values, as provided by the IUPAC facility, although they are 
representative of ``the chemical materials encountered in the laboratory'' not necessarily 
of ``the most abundant natural material'' (Anders \& Grevesse 1989; see also Lodders 2003).

\section{Standard Solar Model}

In order to investigate the change of the solar surface composition induced by different
physical processes and to derive the absolute compositions of the solar photosphere and of 
the solar system, we make use of a Standard Solar Model. We have simulated the evolution of 
the Sun, from the pre-Main Sequence to the present day, by using the FRANEC evolutionary 
code (Chieffi, Limongi \& Straniero 1998).

\subsection{Input physics}

An extended nuclear network, including all the elements from H to U, has been adopted. For 
each element, the evolution of all stable and some long-lived isotopes have been explicitly 
followed (286 isotopes in all). Reaction rates of relevant strong interactions are generally 
taken from the NACRE compilation (Angulo et al. 1999) with few important exception. The 
$\element[][14]{N}(\element[][]{p},\gamma)\element[][16]{O}$ rate is computed according to 
Formicola et al. (2004), while the $\element[][]{D}(\element[][]{p},\gamma)\element[][3]{He}$ 
is derived from Casella et al. (2002). $\beta$-decay rates are taken from Oda et al. (1994) 
for $T> 10^7$ K, while for lower temperature the terrestrial values are assumed. The only 
exception is the $\beta^+$ rate of $\element[][26]{Al}\rightarrow \element[][26]{Mg}$, which 
is derived from Coc et al. (2000). 
The cross section for the $e$-capture on $\element[][7]{Be}$ has been derived from Caughlan 
\& Fowler (1988).

Electron screenings are evaluated according to Dewitt, Graboske \& Cooper (1973) and Graboske 
et al. (1973), for the weak and intermediate regimes, and to Itoh, Totsuji \& Ichimaru (1977) 
and Itoh et al. (1979), for the strong one.
 
Microscopic diffusion of each isotopes is taken into account by inverting the coupled set 
of Burgers equations (Thoul, Bahcall \& Loeb, 1994). In the evaluation of the Coulomb 
logarithm, full ionization has been assumed. This approach, which may be considered as 
``standard'' in the computation of SSMs (see Bahcall, Serenelli \& Pinsonneault 2004, and 
references therein), presents some limitations. Indeed, the ionization degree of the various 
species should be taken into account when computing the diffusion coefficients in the most 
external layers of the star. In practice, the larger the ionization, the larger the Coulomb 
cross section and, in turn, the smaller the diffusion coefficient. 
Since the Sun presents a rather deep external convective envelope, the degree of ionization 
of the various chemical species should be carefully evaluated at the bottom of this zone. 
Another process often neglected is the acceleration caused by the net transfer of momentum 
from the outcoming photon flux to nuclei and electrons. This process acts opposite to the 
gravitational settling on heavy elements. The effect of all these processes on SSMs have 
been properly studied by Turcotte et al. (1998). They computed the fraction of momentum 
transported by radiation that is absorbed by the different chemical species. They also took 
into account the partial ionization in the computation of diffusion coefficients. According 
to their results, the radiative acceleration largely compensates the increase of the diffusion 
efficiency due to partial ionization of the various isotopes. The combined effect of these 
two processes is a variation of the average diffusion efficiency of about 10\% with respect 
the one obtained by neglecting the radiative acceleration and assuming fully ionized matter. 
Turcotte and co-workers also showed that the Rosseland mean opacity calculated by 
interpolating on tables of various $Z$, but the same heavy elements distribution, can not 
be distinguished from that properly computed by taking into account the variation of the 
relative abundances due to all these processes. Comforted by the results of Turcotte et al. 
(1998), in the present work we do not account for these ``non-standard'' effects, even if a 
more precise evaluation of the absolute solar composition would require their proper inclusion 
in the computation of SSMs. In particular, since the degree of ionization changes from one 
nucleus to another, it should be carefully verified that these processes, whose efficiency 
depends on the effective charge, do not alter the abundance ratios of the 34 elements for 
which a very good agreement between the astronomical and the cosmochemical scales have been 
found. We will further comment this point in Section \ref{sec41}.

The borders of the convective unstable zones have been evaluated by means of the Schwartzschildt 
criterion. In these layers, the temperature gradient is calculated by means of the mixing 
length theory (Cox \& Giuli, 1968). As usual, the ratio of the mixing length and the pressure 
scale height is one of the three free parameters of the SSM (see the next section).

The equation of state (EOS) for $T>10^6$ K is an improved version of the one described by 
Straniero (1988) (see Prada Moroni \& Straniero 2002). At lower temperature, a Saha equation 
has been used to derive the population of the various ionization states and of the fractions 
of $\element[][]{H}$ molecules ($\element[][]{H}_2$ and $\element[][]{H}^-$). More accurate 
equations of state, such as the OPAL (Rogers, Swenson \& Iglesias 1996) and the MHD (Mihalas 
et al. 1990, and references therein), may provide a better description of the thermodynamical 
quantities needed to calculate a SSM, but they do not cover the full range of temperature and 
density experienced by the core of off-Main Sequence stars. For this reason we prefer our EOS 
to those commonly adopted by the solar community. A comparison of the SSM computed with our 
EOS and the one obtained by adopting the OPAL EOS is discussed in the next section. 

We have repeated the calculation of the SSM by changing the adopted compilation of relative 
solar abundances, namely: 
\begin{enumerate}
\item {\bf AG89}: Anders \& Grevesse (1989);
\item {\bf GN93}: Grevesse \& Noels (1993);
\item {\bf GS98}: Grevesse \& Sauval (1998);
\item {\bf Lo03}: Lodders (2003).
\end{enumerate}

The main differences among these compilations concern the abundances of carbon, nitrogen,
oxygen, neon and argon. We choose these four compilations because they are considered 
fundamental steps toward our comprehension of the cosmic composition. 
AG89 mixture is a complete and systematic analysis of chemical abundances of the various 
solar system components and it is widely considered the reference framework for studies of 
chemical evolution of the Galaxy. The GN93 compilation has been extensively used in the past 
to compute stellar models (Straniero, Chieffi \& Limongi 1997; Bono et al. 1997; Salaris, 
degl'Innocenti \& Weiss 1997; Pols et al. 1998; Girardi \& Bertelli 1998; Dominguez et al. 
1999; Charbonnel et al. 1999; Bono et al. 2000; Maeder \& Meynet 2001), mainly because the 
opacity tables provided by the OPAL group (Rogers \& Iglesias 1992; Iglesias \& Rogers 1996) 
were originally computed assuming this distribution of heavy elements. The GS98 compilation 
represents the last upgrade of the series started by Anders \& Grevesse (1989) and it has 
been adopted in computing the most recent large database of stellar evolutionary tracks and 
isochrones, ({\it e.g.} Kim et al. 2002; Pietrinferni et al. 2004). Finally, the Lo03 
distribution of heavy elements is representative of the most recent analysis of carbon, 
nitrogen, oxygen, neon and argon abundances as derived by means of 3D-atmospheric models 
(Asplund et al. 2000, 2004; Asplund 2000; Allende Prieto, Lambert \& Asplund 2001, 2002).

For each set of solar abundances, tables of radiative opacity have been derived by means
of the web facility provided by the OPAL group 
(http://www-phys.llnl.gov/Research/OPAL/opal.html). These opacity tables extend down to 
$\log T=3.75$ and are suitable for the calculation of Main Sequence models of a 1 M$_\odot$ 
star. However, additional low temperature opacities are needed to calculate the cooler 
pre-Main Sequence models. In this case, we use the tables provided by Alexander \& Ferguson 
(1994) for a fixed composition (GN93). Note that, since the use of these low temperature 
opacities is limited to pre-Main Sequence models, it does not significantly affect our 
prediction concerning the composition of the present Sun. The opacity change caused by the 
variation of the internal chemical composition, as due to nuclear burning, convective mixing 
and microscopic diffusion, has been taken into account by interpolating between tables with 
different $Z$ and $Y$ (for a description of the numerical procedure adopted to interpolate 
opacity tables see next subsection). 

\subsection{SSM check}

For each set of relative abundances, we vary the initial metallicity ($Z_\mathrm{in}$), 
the helium mass fraction ($Y_\mathrm{in}$) and the value of the $\alpha$ parameter (the 
ratio between mixing length and pressure scale height), until the best reproduction of the 
present-day values of the solar radius ($R_\odot$), luminosity ($L_\odot$) and $(Z/X)_\odot$ 
is obtained. The initial composition ($n_i,\ \mathrm{i=1,\dots ,83}$) is calculated by 
inverting Eq.~(\ref{eq:1}), where $n(\mathrm{H})=(1-Y_\mathrm{in}-Z_\mathrm{in})$, while 
$\varepsilon(\mathrm{el})$ are taken from one of the adopted compilations of relative 
abundances. 

For the solar luminosity, radius, mass and age we adopt $L_\odot=3.844\cdot 10^{33}$ 
erg s$^{-1}$, $R_\odot=6.951\cdot 10^{10}$ cm, $M_\odot= 1.989\cdot 10^{33}$ g and
$t=4.57\cdot 10^9$ yr, respectively. $(Z/X)_\odot$ is obviously different for the four 
set of solar relative abundances, namely: 0.02668 (AG89), 0.02439 (GN93), 0.02292 (GS98) 
and 0.01763 (Lo03). The mass is maintained constant.

In order to investigate the reliability of our input physics we have performed several 
calculations by varying both the interpolation on the tables of opacity coefficients and 
the equation of state and comparing the resulting sound-speed profile with that inferred 
from the GOLF+MDI data (Lazrek et al. 1997; Kosovichev et al. 1997).

The best interpolation method can not be easily fixed according to mathematical or physical 
arguments. Then, we have directly evaluated the influence of this interpolation on the SSM. 
The different numerical schemes we have considered are listed below: 
\begin{itemize}
\item K0: cubic interpolation of 
   $\kappa$ in $R=\rho\cdot 10^{-6}/T$, $T$ and $Y$ and linear interpolation in $Z$;
\item K1: 
   OPACGN93.F routine provided by the OPAL group. Such a 
   routine performs an ``interpolation between overlapping quadratics'' in order 
   to obtain a smoothed opacity;
\item K2: cubic 
   interpolation of $\log(\kappa)$ in $\log(R)$ and $T$ and a linear interpolation 
   of $\kappa$ in $Y$ and $Z$;
\item K3: cubic 
   interpolation of $\log(\kappa)$ in $R$, $T$ and $Y$ and a linear 
   interpolation of $\kappa$ in $Z$.
\end{itemize}
\begin{figure}
   \centering
   \includegraphics[width=\columnwidth]{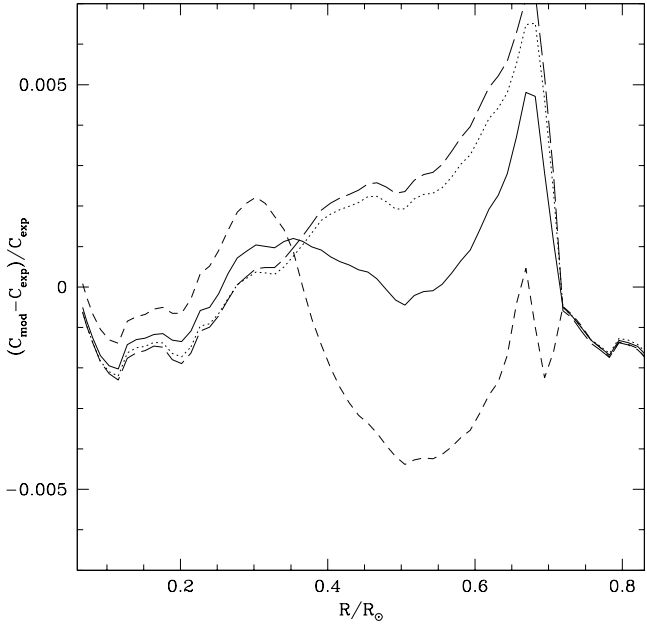}
   \caption{Sound-speed difference between GOLF+MDI data ($\mathrm{C_{exp}}$) and various 
            theoretical models ($\mathrm{C_{mod}}$):
            K0 ({\it solid line}), K1 ({\it long-dashed line}), K2 ({\it dotted line}),
            K3 ({\it dashed line}). For details see text. 
           }
   \label{fig1}
\end{figure}
\begin{table}[h]
\caption{Selected quantities for SSMs obtained by adopting various interpolation schemes for 
         the calculation of opacity coefficients (models from K0 to K3) and by using the most 
         recent OPAL EOS (EOS\_2005): 
         the value of the radius at the base of the convective envelope ($R_\mathrm{CE}$), 
         the mass extension of the convective envelope ($\Delta M_\mathrm{CE}$), 
         the value of the mixing length parameter $\alpha$, 
         the photospheric mass fraction abundance of hydrogen ($X$), helium ($Y$ )and 
         metallicity ($Z$) and their initial values.
        }
        \label{table1}
\centering
\scriptsize
\begin{tabular}{lccccc}
\hline\hline
                                 & K0    & K1    & K2    & K3    & EOS\_2005    \\
\hline
$R_\mathrm{CE}/R_\odot$         & 0.71420 & 0.71755 & 0.71670 & 0.70523 & 0.71431 \\
$\Delta M_\mathrm{CE}/M_\odot$  & 0.02371 & 0.02257 & 0.02284 & 0.02647 & 0.02376 \\
$\alpha$                        & 2.25935 & 2.26263 & 2.27479 & 2.36326 & 1.86171 \\
$X_\odot$                       & 0.73528 & 0.73730 & 0.73751 & 0.72935 & 0.73835 \\
$Y_\odot$                       & 0.24678 & 0.24472 & 0.24450 & 0.25286 & 0.24365 \\
$Z_\odot$                       & 0.01793 & 0.01798 & 0.01799 & 0.01779 & 0.01801 \\
$X_\mathrm{in}$                 & 0.70713 & 0.70322 & 0.70361 & 0.69669 & 0.70581 \\
$Y_\mathrm{in}$                 & 0.27834 & 0.27675 & 0.27636 & 0.28363 & 0.27421 \\
$Z_\mathrm{in}$                 & 0.01994 & 0.02003 & 0.02003 & 0.01968 & 0.01998 \\
\hline
\end{tabular}
\end{table}

The results are summarized in Fig.~\ref{fig1}, where we report the difference of the 
sound-speed inferred from the GOLF+MDI data (Lazrek et al. 1997; Kosovichev et al. 1997) 
and the various test-models. In Tab.~\ref{table1} we list some relevant quantities for the 
same models in Fig.~\ref{fig1}. All these SSM where computed by using the GN93 compilation 
of solar abundances. The position of the inner border of the solar convective envelope 
changes by changing the interpolation scheme. For the particular set of SSM in 
Tab.~\ref{table1}, the variation of $R_\mathrm{CE}$ is one order of magnitude larger than 
the statistical error of the helioseismic data ($\Delta R_\mathrm{CE}=0.001 R_\odot$ - see 
Basu \& Antia 2004, and references therein).
The run of the sound-speed is particularly sensitive to the adopted opacity interpolation 
in the intermediate region, up to the base of the convective envelope ($0.3\le R/R_\odot\le 
0.72$). In the following we will adopt the K0 interpolation scheme.

\begin{figure}
   \centering
   \includegraphics[width=\columnwidth]{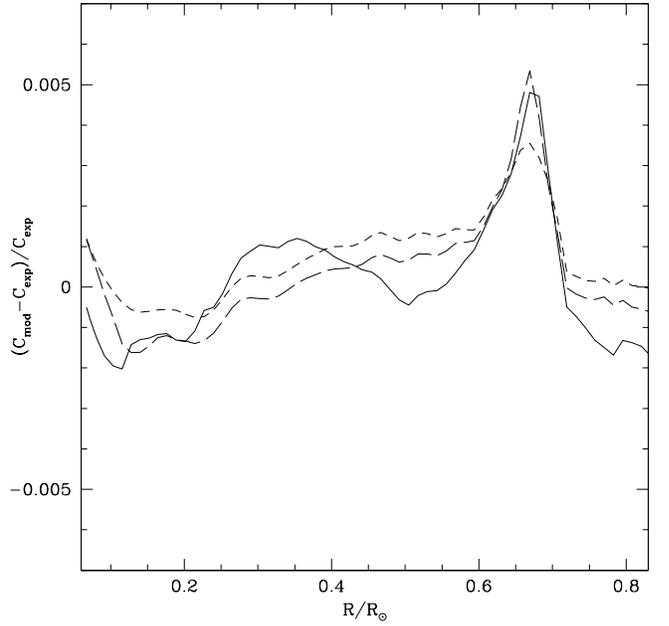}
   \caption{Sound-speed difference between GOLF+MDI data ($\mathrm{C_{exp}}$) and two 
            models ($\mathrm{C_{mod}}$) obtained by 
            adopting different equation of state: our EOS ({\it solid line}) and 
            EOS\_2005 ({\it long-dashed line}). The dashed line refers to the SSM 
            of Bahcall, Basu \& Pinsonneault (1998).
           }
   \label{fig2}
\end{figure}
To check the reliability of the adopted EOS, we have computed an additional SSM by using the 
K0 interpolation and the OPAL EOS\_2005 (http://www-phys.llnl.gov/Research/OPAL/EOS\_2005/). 
The original routine provided by OPAL has been used to interpolate on the EOS tables. The 
sound-speed difference between the experimental data and this SSM is shown in Fig.~\ref{fig2}. 
In the same figure, we also report the same quantity for the K0 model of Fig.~\ref{fig1} and 
for the SSM by Bahcall, Basu \& Pinsonneault (1998). 
All the 3 models provide a good reproduction of the measured sound-speed, the deviation being 
confined within $\pm 0.002$, with the exception of the well known discrepancy at the base of 
the external convective zone. As expected, in the most external zone, the EOS\_2005 provides 
a better evaluation of the thermodynamical quantities as compared to the Saha equation. Note 
that the SSMs computed with the Saha EOS require a rather large value of $\alpha$, {\it i.e.} 
the mixing length parameter. As a whole, the model obtained with our EOS (K0) provides a 
reasonable reproduction of the present Sun, at least comparable to the one obtained with the 
OPAL EOS (see Tab.~\ref{table1}).

In Tab.~\ref{table2} we summarize the main features of the SSMs computed by adopting the four 
compilation of solar abundances described in the previous section. The most important 
difference is the variation of the initial (pre-solar) metallicity ($Z_\mathrm{in}$), which 
decreases from $Z=0.02177$ (AG89) to 0.01487 (Lo03). The most recent compilation of solar 
abundances also implies a lower initial abundance of He ($Y_\mathrm{in}$ in Tab.~\ref{table2}), 
namely $Y_\mathrm{in}=$0.2776 (AG89) and 0.2674 (Lo03). 

The simultaneous reduction of $Z_\mathrm{in}$ and $Y_\mathrm{in}$ leaves almost unchanged the 
ratio $\Delta Y/ \Delta Z$. Indeed, by taking $Y=0.245$ for the cosmological He (Spergel et 
al. 2003), one gets $\Delta Y/ \Delta Z \sim 1.5$ for both AG89 and Lo03. This value is 
definitely lower then that derived from $\ion{H}{ii}$ regions and planetary nebulae (see 
Esteban \& Peimbert 1995). 

Nevertheless, these differences in $Z$ and $Y$ affect the opacity and, in turn, the internal 
temperature profile and the stellar radius. In this context, the reduction of the metallicity 
is partially counterbalanced by the reduction of the $\element[][]{He}$ content. The combined 
effect is a net reduction of the opacity, when passing from AG89 to Lo03. This reduction is 
about 7\% near the base of the convective envelope (Bahcall, Serenelli \& Pinsonneault 2004). 
As a consequence, the external convective zone shrinks (see $R_\mathrm{CE}$ in 
Tab.~\ref{table2}). Note that in the Lo03 model, the location of the inner border of the 
convective envelope is definitely larger than the one derived from the observed frequencies 
of solar oscillations, namely $R_\mathrm{CE}=0.7295$ to be compared with $0.713\pm 0.001$ 
(Basu \& Antia 1997). In addition, a lower $\alpha$ is required to reproduce the observed 
solar radius. This occurrence may affect the properties of stellar models that, as usual, 
adopt a solar calibrated mixing length parameter. We will briefly discuss this point in the 
final section.
\begin{table}[h]
\caption{Selected quantities for our best SSMs computed by adopting different abundance
         compilations: 
         the value of the radius at the base of the convective envelope ($R_\mathrm{CE}$), 
         the mass extension of the convective envelope ($\Delta M_\mathrm{CE}$), 
         the temperature ($T_\mathrm{CE}$) and density ($\rho_\mathrm{CE}$) at the base 
         of the convective envelope, 
         the value of the mixing length parameter $\alpha$, 
         the central value of temperature (T$_\mathrm{c}$) and density ($\rho_\mathrm{c}$), 
         hydrogen ($X_\mathrm{c}$) and helium ($Y_\mathrm{c}$) abundances at the center 
         (by mass fraction), 
         the current-age photospheric $(Z/X)$, $X$, $Y$ and $Z$ and 
         their initial values (see text).
        }
        \label{table2}
\centering
\begin{tabular}{lcccc}
\hline\hline
                                     & AG89    & GN93    & GS98    & Lo03   \\
\hline
$R_\mathrm{CE}/R_\odot$             & 0.71110 & 0.71426 & 0.71621 & 0.72949 \\
$\Delta M_\mathrm{CE}/M_\odot$      & 0.02459 & 0.02367 & 0.02316 & 0.01940 \\
$T_\mathrm{CE}$ (in 10$^6$ K)       & 2.22662 & 2.18691 & 2.16374 & 2.01571 \\
$\rho_\mathrm{CE}$ (in g cm$^{-3}$) & 0.19015 & 0.18349 & 0.17983 & 0.15255 \\
$\alpha$                            & 2.27016 & 2.25772 & 2.25112 & 2.14243 \\
T$_\mathrm{c}$ (in 10$^6$ K)        & 15.7591 & 15.7065 & 15.6846 & 15.5818 \\
$\rho_\mathrm{c}$ (in g cm$^{-3}$)  & 155.008 & 154.380 & 154.169 & 152.714 \\
X$_\mathrm{c}$                      & 0.32948 & 0.33419 & 0.33636 & 0.34680 \\
Y$_\mathrm{c}$                      & 0.64675 & 0.76554 & 0.64297 & 0.63698 \\
$(Z/X)_\odot$                       & 0.02669 & 0.02439 & 0.02292 & 0.01762 \\
$X_\odot$                           & 0.73103 & 0.73542 & 0.73757 & 0.74792 \\
$Y_\odot$                           & 0.24940 & 0.24664 & 0.24553 & 0.23890 \\
$Z_\odot$                           & 0.01951 & 0.01793 & 0.01690 & 0.01381 \\
$(Z/X)_\mathrm{in}$                 & 0.03107 & 0.02841 & 0.02673 & 0.02074 \\
$X_\mathrm{in}$                     & 0.69732 & 0.70175 & 0.70383 & 0.71267 \\
$Y_\mathrm{in}$                     & 0.28102 & 0.27831 & 0.27736 & 0.27256 \\
$Z_\mathrm{in}$                     & 0.02166 & 0.01994 & 0.01881 & 0.01478 \\
\hline
\end{tabular}
\end{table}

\section{Proto-solar {\it vs.} Solar Abundances}

In this section we illustrate the procedure followed to evaluate the mass fractions 
of isotopes and elements at the photosphere of the present Sun as well as in the 
early-solar-system material. 

First of all let us assume that:
\begin{equation}
\left(\frac{n_k}{n_j}\right)_\odot=\left(\frac{n_k}{n_j}\right)_\mathrm{CI} \\ j,k=3,\dots ,83
\label{eq:5}
\end{equation}
\noindent
where the suffixes ``$\odot$'' and ``$\mathrm{CI}$'' refer to the solar photosphere and the
pristine material (CI carbonaceous chondrites), respectively. This is a generalization of the 
assumption already made to translate the cosmochemical into the astronomical scale (see 
Eq.~(\ref{eq:3}), where $j=14$, the $\element[][]{Si}$ atomic number). Eq.~(\ref{eq:5}) has 
been proved for most of the elements whose abundances can be derived from both meteorites 
and the solar photosphere, with the unique clear exception of lithium. However, since the 
$\element[][]{Li}$ abundance is small compared to the total metallicity, its deviation from 
this general rule does not affect our estimation of the absolute abundances of other elements. 
Thus, let us extend the validity of Eq.~(\ref{eq:5}) to all the elements, $\element[][]{H}$ 
and $\element[][]{He}$ excluded.

Since $(x_k/x_j)=(n_k/n_j)\cdot(A_k/A_j)$ \footnote {$x_j$ is the mass fraction of the 
$j^{th}$ element and $A_j$ is the corresponding average atomic mass. For each element, 
this average can be calculated by means of the terrestrial isotopic ratios provided by 
IUPAC.}, the previous relation can be re-written as:
\begin{equation}
\left(\frac{x_k}{x_j}\right)_\odot=\left(\frac{x_k}{x_j}\right)_\mathrm{CI}
\label{eq:6}
\end{equation}
\noindent

Here we are assuming that the isotopic ratios remains (almost) unchanged over the solar 
system lifetime, so that 
$\left(\frac{A_k}{A_j}\right)_\odot=\left(\frac{A_k}{A_j}\right)_\mathrm{CI}$. For a few 
elements with radioactive isotopes whose decay time is comparable to the solar age (see 
next section), this assumption is not correct. Also in this case, since their abundances 
are small compared to the total metallicity and their impact on the SSM is negligible, 
we can correct their early solar system abundances, without affecting the evaluation of 
the abundances of the other elements. Then, by summing over the index $k$, we obtain 
$Z/x_j$ \footnote{Here $Z$ represents the total metallicity.} and dividing Eq.(~\ref{eq:6}) 
by this quantity, it follows:
\begin{equation}
\left(\frac{x_k}Z\right)_\odot=\left(\frac{x_k}Z\right)_\mathrm{CI} \\ k=3,\dots ,83
\label{eq:7}
\end{equation}
\noindent
In other words, the mass fraction of a given element in the solar photosphere and the 
corresponding one in the early solar system material scale as the ratio 
$Z_\odot/Z_\mathrm{in}$. Hence, we can invert Eq.~(\ref{eq:1}) to find the following set 
of equations:
\begin{equation}
\frac{x_k}{X} = 10^{\varepsilon(k)-12}\frac {A_k}{A_H} \\ k=3,\dots ,83
\label{eq:8}
\end{equation}
\noindent
where $X$ is the hydrogen mass fraction. If $X$ would be known, the mass fractions of all 
the elements could be derived from these equations, but, unfortunately, it can not be 
directly measured at the solar photosphere. Note that, since we have used the astronomical 
scale, this equation can only be applied to the photospheric abundances. A more general 
relation, valid for any scaled solar composition, can be determined by summing all the 
Eqs.~(\ref{eq:8}), thus obtaining: 
\begin{equation}
\left(\frac Z X\right)_\odot = \sum_{k=3}^{83} {10^{\varepsilon(k)-12} \frac{A_k}{A_H}}
\label{eq:9}
\end{equation}
\noindent
Then, dividing Eqs.~(\ref{eq:8}) by Eq.~(\ref{eq:9}) we find:
\begin{equation}
\frac {x_k} Z = \frac{10^{\varepsilon(k)-12} \frac{A_k}{A_H}}{\sum_{j=3}^{83}
{10^{\varepsilon(j)-12} \frac{A_j}{A_H}}}=10^{\varepsilon(k)-12} \frac{A_k}{A_H} \left(\frac Z X\right)_\odot^{-1} \\ k=3,\dots ,83
\label{eq:10}
\end{equation}

Remembering our basic assumption, as expressed in Eq.~(\ref{eq:7}), this formula can be used 
to calculate the early solar system composition, once the value of $Z_\mathrm{in}$ is known. 
The same formula can be used to derive the absolute abundances of any star having a 
scaled-solar composition, once its metallicity ($Z$) is given. 
\begin{figure*}
   \centering
   \includegraphics[width=\textwidth]{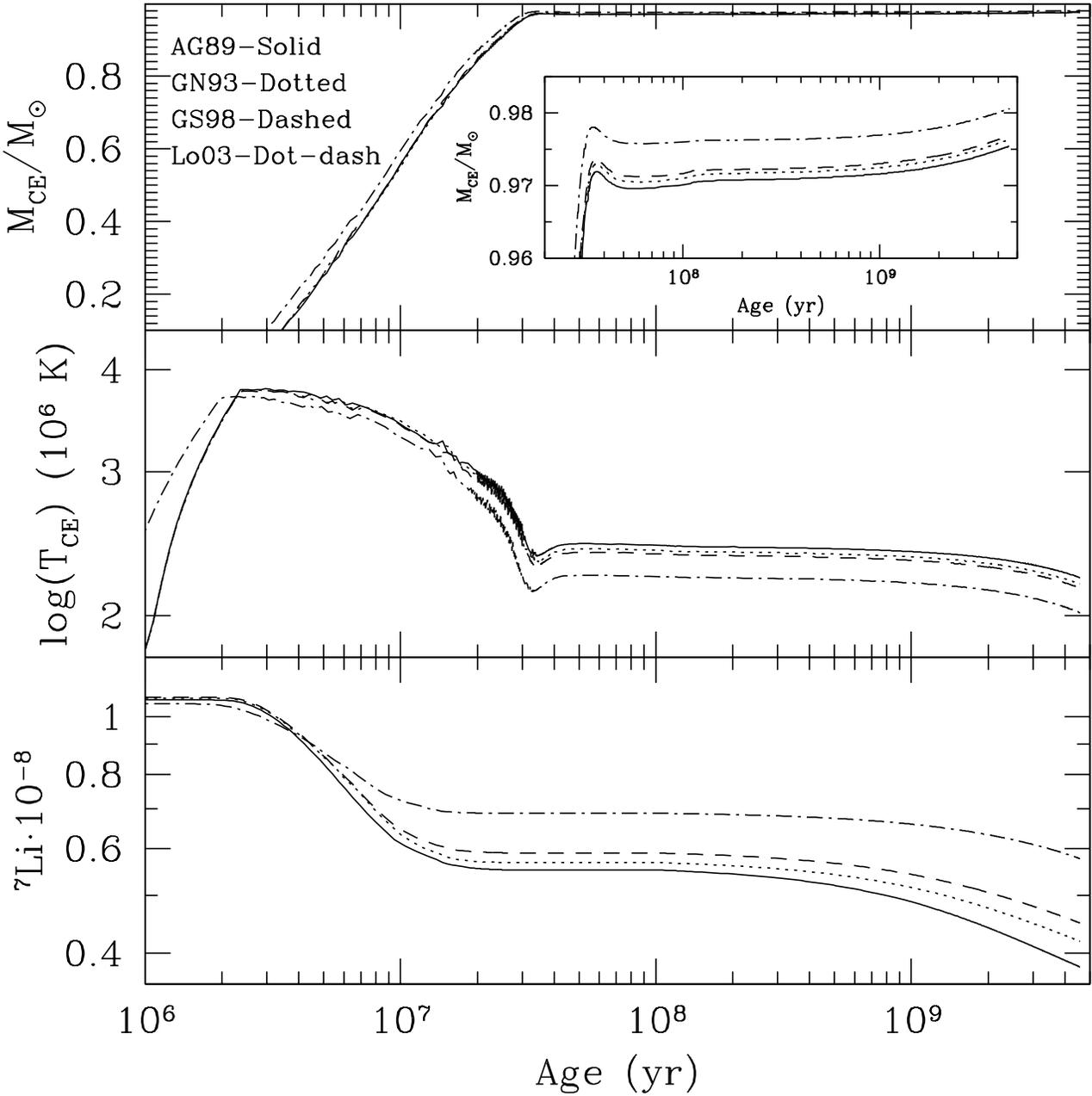}
   \caption{Evolution of:
            mass coordinate (in solar unit) of the inner border of the convective envelope ({\it upper panel});
            temperature at the base of the convective envelope ({\it middle panel});
            lithium abundance by mass fraction ({\it lower panel}).
           }
   \label{fig3}
\end{figure*}

In practice, we start with a tentative set of $Z_\mathrm{in}$, $Y_\mathrm{in}$ and $\alpha$.
Then, by means of the Eqs.~(\ref{eq:10}) and by taking terrestrial ratios among isotopes of 
the same element (but applying appropriate corrections to the radioactive isotopes), we 
calculate the initial mass fraction of all the 286 isotopes and we compute a first 
evolutionary sequence up to the solar age. Then, the initial values of $Z$, $Y$ and $\alpha$ 
are corrected and the procedure is iterated, until a good reproduction of $R_\odot$, $L_\odot$ 
and $(Z/X)_\odot$ is simultaneously obtained\footnote{We stop the iterations when 
$\delta R/R_\odot$, $\delta L/L_\odot$ and $\delta (Z/X)/(Z/X)_\odot$ become smaller then 
$10^{-5}$.}. Note that, for the initial composition, we generally choose meteoritic (CI) 
relative abundances, if available. For this reason, we obtain $(Z/X)_\odot$ values that 
are slightly different from those quoted in the original papers reporting the four solar 
abundance compilations, where somewhat different choices are adopted.
\addtocounter{table}{1}
\onllongtabL{3}{
\begin{landscape}
\tiny
\begin{longtable}{c|c|c|ccc|ccc|ccc|ccc}
\caption{\label{table3} From left to right, mass fraction abundances for all the considered isotopes 
            of the proto-solar nebula ($x_\mathrm{in}$) and at the solar photosphere 
            ($x_\odot$). The depletion efficiency $\delta$ is also reported. 
            The initial and final values of the metallicity and the corresponding depletion 
            efficiency are reported in the last row (see text).}\\
\hline\hline
  \multicolumn{3}{c|}{ }  &
  \multicolumn{3}{c|}{AG89}  &
  \multicolumn{3}{c|}{GN93}  &
  \multicolumn{3}{c|}{GS98}  &
  \multicolumn{3}{c}{Lo03}    \\
  \cline{1-3}\cline{4-6}\cline{7-9}\cline{10-12}\cline{13-15}
  El & Z & A &
  $x_\mathrm{in}$ & $x_{\odot}$ & $\delta$ &
  $x_\mathrm{in}$ & $x_{\odot}$ & $\delta$ &
  $x_\mathrm{in}$ & $x_{\odot}$ & $\delta$ &
  $x_\mathrm{in}$ & $x_{\odot}$ & $\delta$  \\
\hline
\endfirsthead
\caption{continued.}\\
\hline\hline
  \multicolumn{3}{c|}{ }  &
  \multicolumn{3}{c|}{AG89}  &
  \multicolumn{3}{c|}{GN93}  &
  \multicolumn{3}{c|}{GS98}  &
  \multicolumn{3}{c}{Lo03}    \\
  \cline{1-3}\cline{4-6}\cline{7-9}\cline{10-12}\cline{13-15}
  El & Z & A &
  $x_\mathrm{in}$ & $x_{\odot}$ & $\delta$ &
  $x_\mathrm{in}$ & $x_{\odot}$ & $\delta$ &
  $x_\mathrm{in}$ & $x_{\odot}$ & $\delta$ &
  $x_\mathrm{in}$ & $x_{\odot}$ & $\delta$  \\
\hline
\endhead
\hline
\endfoot
H    &  1 &   1.007825 & 0.6973     & 0.7310     &   -4.84 & 0.7017     & 0.7354     &   -4.80& 0.7038     & 0.7376     &   -4.80 & 0.7127     & 0.7479     &   -4.95 \\
     &    &   2.014102 & 2.3709E-05 & 2.0297E-17 &  100.00 & 2.3860E-05 & 2.0738E-17 &  100.00& 2.3930E-05 & 2.1005E-17 &  100.00 & 1.3826E-05 & 2.3034E-17 &  100.00 \\
He   &  2 &   3.016029 & 3.9904E-05 & 6.9378E-05 &  -73.86 & 3.9520E-05 & 6.9141E-05 &  -74.95& 3.9385E-05 & 6.9056E-05 &  -75.34 & 4.5235E-05 & 6.0081E-05 &  -32.82 \\
     &    &   4.002603 & 0.2810     & 0.2494     &   11.24 & 0.2783     & 0.2466     &   11.39& 0.2773     & 0.2455     &   11.49 & 0.2725     & 0.2388     &   12.36 \\
Li   &  3 &   6.015123 & 7.4216E-10 & 2.4842E-16 &  100.00 & 7.4719E-10 & 2.3647E-16 &  100.00& 7.5013E-10 & 2.4387E-16 &  100.00 & 7.2349E-10 & 1.1082E-13 &   99.98 \\
     &    &   7.016005 & 1.0676E-08 & 3.7929E-09 &   64.47 & 1.0749E-08 & 4.1896E-09 &   61.02& 1.0791E-08 & 4.4981E-09 &   58.32 & 1.0276E-08 & 5.6411E-09 &   45.10 \\
Be   &  4 &   9.012182 & 1.9100E-10 & 1.6900E-10 &   11.52 & 1.9229E-10 & 1.6988E-10 &   11.65& 1.9305E-10 & 1.7038E-10 &   11.74 & 1.9268E-10 & 1.6847E-10 &   12.56 \\
B    &  5 &  10.012937 & 1.2179E-09 & 1.0918E-09 &   10.35 & 9.9665E-10 & 8.9224E-10 &   10.48& 1.0006E-09 & 8.9492E-10 &   10.56 & 9.9465E-10 & 8.8207E-10 &   11.32 \\
\end{longtable}
\end{landscape}
}

\addtocounter{table}{1}
\onllongtabL{4}{
\begin{landscape}
\tiny
\begin{longtable}{c|cccc|cccc|cccc|cccc}
\caption{\label{table4} For each adopted solar distribution, the initial ($x_\mathrm{in}$) and  
            the present-day ($x_\odot$) mass fractions and the corresponding values 
            of the $P$ and $R$ parameters.}\\
\hline\hline
                   &
  \multicolumn{4}{c|}{AG89}  &
  \multicolumn{4}{c|}{GN93}  &
  \multicolumn{4}{c|}{GS98}  &
  \multicolumn{4}{c }{Lo03}  \\
  \cline{2-5}\cline{6-9}\cline{10-13}\cline{14-17}
 El. &
  $x_\mathrm{in}$ & $x_{\odot}$ & $P$ & $R$ &
  $x_\mathrm{in}$ & $x_{\odot}$ & $P$ & $R$ &
  $x_\mathrm{in}$ & $x_{\odot}$ & $P$ & $R$ &
  $x_\mathrm{in}$ & $x_{\odot}$ & $P$ & $R$ \\
\hline
\endfirsthead
\caption{continued.}\\
\hline\hline
                   &
  \multicolumn{4}{c|}{AG89}  &
  \multicolumn{4}{c|}{GN93}  &
  \multicolumn{4}{c|}{GS98}  &
  \multicolumn{4}{c }{Lo03}  \\
  \cline{2-5}\cline{6-9}\cline{10-13}\cline{14-17}
 El. &
  $x_\mathrm{in}$ & $x_{\odot}$ & $P$ & $R$ &
  $x_\mathrm{in}$ & $x_{\odot}$ & $P$ & $R$ &
  $x_\mathrm{in}$ & $x_{\odot}$ & $P$ & $R$ &
  $x_\mathrm{in}$ & $x_{\odot}$ & $P$ & $R$ \\
\hline
\endhead
\hline
\endfoot
H  & 0.6973     & 0.7310     & 0.862 & 1.616 & 0.7018     & 0.7354     & 0.861 & 1.616 & 0.7038     & 0.7376     & 0.861 & 1.627 & 0.7127     & 0.7479     & 0.853 & 1.611 \\
He & 0.2810     & 0.2495     & 1.018 & 1.544 & 0.2783     & 0.2466     & 1.018 & 1.544 & 0.2774     & 0.2455     & 1.019 & 1.553 & 0.2726     & 0.2389     & 1.021 & 1.532 \\
\hline
Li & 1.1419E-08 & 3.7929E-09 & 2.750 & 1.112 & 1.1496E-08 & 4.1896E-09 & 2.503 & 1.153 & 1.1541E-08 & 4.4981E-09 & 2.339 & 1.193 & 1.0999E-08 & 5.6412E-09 & 1.764 & 1.295 \\
Be & 1.9100E-10 & 1.6900E-10 & 1.021 & 1.543 & 1.9229E-10 & 1.6988E-10 & 1.022 & 1.542 & 1.9305E-10 & 1.7038E-10 & 1.022 & 1.552 & 1.9268E-10 & 1.6847E-10 & 1.023 & 1.531 \\
B  & 6.6079E-09 & 5.8904E-09 & 1.014 & 1.546 & 5.4075E-09 & 4.8133E-09 & 1.014 & 1.545 & 5.4287E-09 & 4.8275E-09 & 1.014 & 1.555 & 5.4188E-09 & 4.7753E-09 & 1.015 & 1.535 \\
C  & 3.5138E-03 & 3.1554E-03 & 1.006 & 1.549 & 3.4571E-03 & 3.1004E-03 & 1.006 & 1.549 & 3.2390E-03 & 2.9023E-03 & 1.006 & 1.559 & 2.4524E-03 & 2.1791E-03 & 1.007 & 1.538 \\
N  & 1.2663E-03 & 1.1387E-03 & 1.005 & 1.550 & 1.0604E-03 & 9.5235E-04 & 1.005 & 1.549 & 9.4879E-04 & 8.5134E-04 & 1.005 & 1.559 & 7.8766E-04 & 7.0101E-04 & 1.005 & 1.539 \\
O  & 1.0972E-02 & 9.8790E-03 & 1.004 & 1.550 & 9.6212E-03 & 8.6518E-03 & 1.004 & 1.550 & 8.8091E-03 & 7.9148E-03 & 1.004 & 1.560 & 6.5179E-03 & 5.8090E-03 & 1.004 & 1.540 \\
\end{longtable}
\end{landscape}
}

In Tab.~\ref{table3} we list the isotopic composition of selected elements (initial and 
present-day) for the four SSMs. In the same table we also report the corresponding depletion 
efficiency $\delta=100\cdot \left(1-X_{\odot}/X_\mathrm{in}\right)$. In Tab.~\ref{table4} 
we report, for the four SSMs, the initial mass fraction, the present-day mass fraction, the 
ratio $P=\left(n(\mathrm{el})/n(\mathrm{Si})\right)_\mathrm{in}/\left(n(\mathrm{el})/n(\mathrm{Si})\right)_{\odot}$
and the $R$ parameter, for the 22 elements whose isotopic composition is reported in 
Tab.~\ref{table3}\footnote{
The full version of Tab.~\ref{table3}, including all the 286 isotopes, and Tab.~\ref{table4}, 
including all the 83 elements, can be found in the electronic version.}. Now, we can check 
the consistency of our basic assumption, that is: the evolution of the solar surface 
composition has occurred preserving the abundance ratios of heavy elements.
\begin{figure*}
   \centering
   \includegraphics[width=\textwidth]{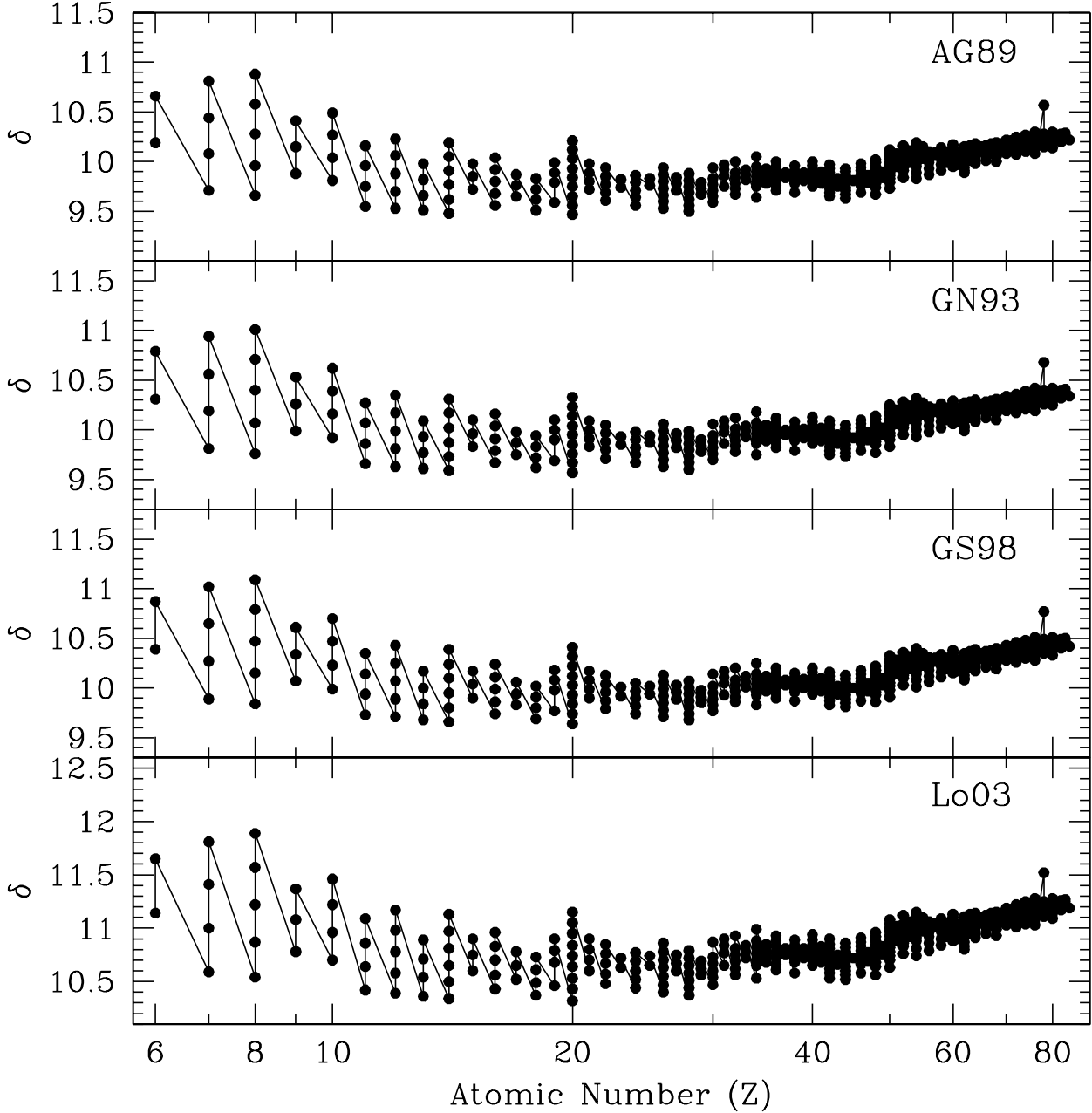}
   \caption{Depletion efficiency $\delta=100\cdot (1-X_{\odot}/X_\mathrm{in})$ as a
            function of the atomic number $Z$ for all the isotopes with $Z\ge 6$.
            Each panel refers to a different abundance compilation, as labeled (see text). 
            We omit the value corresponding to $\beta$-unstable isotopes, except 
            that of the $\element[][190]{Pt}$ (Z=78).
           }
   \label{fig4}
\end{figure*}

\subsection {$Z\ge 6$}\label{sec41}

Owing to the rather fast mixing induced by convection, the composition within the most 
external 30\% of the solar radius is maintained homogeneous. Thus, the photospheric 
composition changes mainly due to processes occurring at the base of the convective 
envelope. Concerning elements with $Z\ge 6$, they are not affected by nuclear burning, 
because the temperature within this convective layer never exceeds few millions of K 
(see the middle panel in Fig.~\ref{fig3}), not enough for $p$-capture to occur on these 
elements. As a consequence the variation of their surface abundances is only determined 
by microscopic diffusion occurring at the base of the convective envelope. 
Let us start discussing the SSM computed by adopting the AG89 mixture. An inspection of 
Tab.~\ref{table3} reveals that the depletion efficiency of the various isotopes ranges 
from 9.7 to 10.7. 

By considering different isotopes of the same element ({\it e.g.} 
$\element[][12]{C}$ and $\element[][13]{C}$ or $\element[][20]{Ne}$ and 
$\element[][22]{Ne}$) it results that the higher the atomic weight, the larger the 
depletion efficiency (see Fig.~\ref{fig4}). A simple argument may explain this feature. 
Heavy nuclei, at the inner border of the solar convective envelope, move downward against 
the buoyancy, which is proportional to their weight $A$, and are decelerated by the 
scattering with the surrounding charge particles, which depends on the atomic number $Z$ 
(we assume full ionization). 
Thus, the diffusion efficiency is expected to roughly scale as $A/Z$. Such an argument also
explains why heavier isotopes of the same element are more efficiently depleted than the 
lighter ones. Note that the minimum depletion occurs around $\element[][40]{Ca}$ (see 
Fig.~\ref{fig4}). Indeed, stable nuclei with $A>40$ are progressively more neutronized,
so that the ratio $A/Z$ becomes systematically larger.

From the data reported in Tab.~\ref{table4}, we obtain ratios of the initial to present-day 
relative abundances very close to 1, the average value being $P=1.003\pm 0.004$. The 
corresponding average value of $R$ is $1.552\pm 0.002$ \footnote{The "errors" on $P$ and $R$ 
represent the maximum semi dispersion.}, in excellent agreement with its experimental 
determination, namely $R=1.556\pm 0.008$ (Palme \& Jones, 2004). As expected, few exceptions 
are represented by elements having unstable isotopes with half life comparable to the age of 
the Sun, namely: $\element[][40]{K}$, $\element[][87]{Rb}$, $\element[][138]{La}$, 
$\element[][147]{Sm}$, $\element[][176]{Lu}$, $\element[][187]{Re}$, $\element[][190]{Pt}$, 
$\element[][232]{Th}$, $\element[][235]{U}$, $\element[][238]{U}$ \footnote{Such elements 
have been excluded from the quoted average of $P$ and $R$.}.

Similar conclusions can be derived for the other SSMs. In particular, the average isotopic 
diffusion efficiency is 9.98, 10.09, 10.18, 10.91 for AG89, GN93, GS98 and Lo03, respectively, 
with a rather small dispersion ($\sim \pm 0.8$). The differences of the average diffusion 
efficiency between various SSMs can be explained simply observing that such a quantity 
depends on $R_\mathrm{CE}^2$. Indeed, the larger $R_\mathrm{CE}$, the larger the spherical 
surface from which heavy elements within the convective envelope are diffused downward. 

If the effects of partial ionization on microscopic diffusion and radiative acceleration 
were simultaneously taken into account, a slightly larger depletion efficiency of metal 
would be obtained. According to Turcotte et al (1998), the increase of the depletion 
efficiency is less than 10\% (see their Fig. 14). Since both these processes depend on the 
degree of ionization that is different from one nucleus to another, the scatter of the 
depletions is expected to increase. On the base of the calculations of Turcotte and co-workers, 
it turns out that the average values of $P$ and $R$ should remain almost unaffected, while 
the dispersions should become about 50\% larger, but in any case, within the experimental 
error bars.

\subsection{Light isotopes ($3<Z<6$)}

All the computed SSMs predict a stronger surface depletion for light elements compared to 
that of the heavy ones. In particular, the ratios $n(\mathrm{light\ el})/n(\mathrm{Si})$ 
is not conserved. Concerning beryllium and boron, such an occurrence can be explained noting 
that the most abundant isotopes, $\element[][9]{Be}$ and $\element[][11]{B}$, have large 
$A/Z$ values. In any case, the initial to final abundance ratio is consistent with the 
corresponding meteoritic to spectroscopic abundance ratio. 

\begin{figure}
   \centering
   \includegraphics[width=\columnwidth]{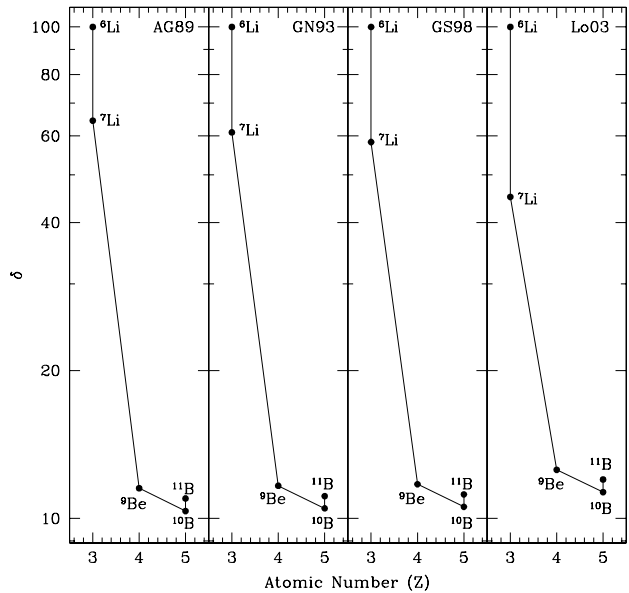}
   \caption{The same as in Fig.~\ref{fig4}, but for light elements ($3\le Z\le 5$).
           }
   \label{fig5}
\end{figure}

On the other hand, the evolution of the solar lithium is one of the most debated problems in 
stellar modeling. During the past 4.5 Gyr, both $\element[][6]{Li}$ and $\element[][7]{Li}$ 
suffered $p$-captures at the base of the convective external layer. Hence, not only 
microscopic diffusion, but also nuclear burning contributed to the modification of the 
photospheric lithium abundance. The major depletion takes place during the pre-Main Sequence, 
when, leaved the Hayashy track, the solar model develops a convective envelope, whose inner 
border progressively recedes in mass (see Fig.~\ref{fig3}). The temperature at this border 
remains almost constant ($T_\mathrm{CE}\sim 3.8\dot 10^6$ K), for about $10^7$ yr, so that 
the $\element[][]{Li}$ consumption is rather efficient during this phase. When the mass of the 
convective envelope is reduced down to $\delta M_\mathrm{CE} \sim 0.4 M_\odot$, 
$T_\mathrm{CE}$ starts to decrease. Then, when the solar age is about $2\cdot 10^7$ yr and 
$\delta M_\mathrm{CE}\sim 0.2 M_\odot$, the temperature becomes too low and the burning of 
lithium ceases. During the following Main Sequence evolution, the temperature at the base 
of the convective envelope always remains too small (in all the four SSMs), so that the 
nuclear burning is inefficient (at least for $\element[][7]{Li}$, the most abundant isotope). 
Hence, for age greater than $10^8$ yr, microscopic diffusion coupled to the convective mixing 
is the dominant physical process for the lithium depletion. As a result, $\element[][6]{Li}$ 
is almost completely destroyed, while the present-day $\element[][7]{Li}$ is 3 to 5 times 
smaller than the initial one (see Fig.~\ref{fig5}). This theoretical result would imply a 
photospheric value of $\varepsilon(\mathrm{Li})$ ranging between 2.86 (AG89) to 3.05 (Lo03), 
to be compared with the corresponding spectroscopic determination, namely $1.1\pm 0.1$ 
(Carlsson et al. 1994). 

This huge discrepancy represents the longstanding ``solar lithium problem'', for which no 
satisfactory explanations have been found up to now. Among the many attempts, it has been 
suggested that a turbulent mixing would be active at the inner border of the solar convective 
envelope. Such a process should be driven by the expected discontinuity in the 
internal-rotational-velocity profile of the Sun. Indeed, according to the observation of 
solar surface oscillations and the relative analysis of the normal modes, it has been derived 
that a latitudinal-differential rotation takes place in the convective envelope, while a 
rigid-body rotation is active in the deep radiative interior (Goode et al. 1991). Then, 
there exists a thin layer, the so called {\it tachocline}, located just below the convective 
envelope, where the external-latitudinal-differential rotation smoothly matches onto the 
internal-uniform rotation. In this case, the resulting mixing does depend only on macroscopic 
properties of the star, as the local gradient of angular velocity, so that it equally affects 
all the chemical species. The effect of this mixing on SSM has been studied by Richard et al. 
(1996) and Brun, Turck-Chi\`eze \& Zahn (1999 - hereinafter BTZ99). In these works 
microscopic diffusion as well as rotational-induced mixing in the radiative zones are 
described by means of a diffusion equation. In particular in BTZ99 the diffusion coefficients 
related to tachocline mixing are computed according to the prescription of Spiegel \& Zahn 
(1992). The exact value of the tachocline-diffusion coefficient depends on the solar angular 
velocity at the base of the convective envelope, the solar differential rotation rate, the 
extension of the tachocline and the horizontal diffusivity. 
\begin{table}
\caption{Some relevant quantities for models computed by including (GN93-R) and 
         not including (GN93) the rotational-induced mixing in the tachocline: 
         the value of the radius at the base of the convective envelope ($R_\mathrm{CE}$),
         the mass extension of the convective envelope ($\Delta M_\mathrm{CE}$), 
         the temperature ($T_\mathrm{CE}$) and density ($\rho_\mathrm{CE}$) at the base 
         of the convective envelope, 
         the central value of temperature (T$_\mathrm{c}$) and density ($\rho_\mathrm{c}$), 
         the value of the mixing length parameter ($\alpha$), 
         the current-age photospheric $(Z/X)$, $X$, $Y$ and $Z$ and 
         their initial values, the depletion efficiency of the total metallicity ($\delta(Z)$),
         the abundances of the light metals lithium, beryllium and boron at the solar photosphere 
         and the corresponding value of the $P$ parameter (see text).
        }
\label{table5}
\centering
\begin{tabular}{lrr}
\hline\hline
                                    & GN93      & GN93-R    \\
\hline
$R_\mathrm{CE}/R_\odot$             & 0.71426   & 0.71702   \\
$\Delta M_\mathrm{CE}/M_\odot$      & 0.02367   & 0.02298   \\
$T_\mathrm{CE}$ (in 10$^6$ K)       & 2.18691   & 2.16688   \\
$\rho_\mathrm{CE}$ (in g cm$^{-3}$) & 0.18349   & 0.17865   \\
T$_\mathrm{c}$ (in 10$^6$ K)        & 15.7065   & 15.6489   \\
$\rho_\mathrm{c}$ (in g cm$^{-3}$)  & 154.380   & 153.804   \\
$\alpha$                            & 2.25772   & 2.21788   \\
$(Z/X)_\odot$                       & 0.02439   & 0.02439   \\
$X_\odot$                           & 0.73542   & 0.72995   \\
$Y_\odot$                           & 0.24664   & 0.25225   \\
$Z_\odot$                           & 0.01793   & 0.01780   \\
$(Z/X)_\mathrm{in}$                 & 0.02841   & 0.02692   \\
$X_\mathrm{in}$                     & 0.70175   & 0.70680   \\
$Y_\mathrm{in}$                     & 0.27831   & 0.27418   \\
$Z_\mathrm{in}$                     & 0.01994   & 0.01902   \\
$\delta(Z)$                         & 10.08     & 6.41      \\
$x(\element[][6]Li)_\odot$          & 2.365e-16 & 3.673e-16 \\
$x(\element[][7]Li)_\odot$          & 4.190e-09 & 2.865e-10 \\
$x(\element[][9]Be)_\odot$          & 1.699e-10 & 1.638e-10 \\
$x(\element[][10]B)_\odot$          & 8.922e-10 & 8.933e-10 \\
$x(\element[][11]B)_\odot$          & 3.921e-09 & 3.741e-09 \\
$P(\mathrm Li)$                     & 2.51      & 36.35     \\
$P(\mathrm Be)$                     & 1.02      & 1.05      \\
$P(\mathrm B)$                      & 1.01      & 1.04      \\
\hline
\end{tabular}
\end{table}

Following the prescription of BTZ99 we have computed an additional SSM (GN93-R), accounting 
for mixing in the tachocline. The diffusion coefficients related to such a mixing have been 
calculated by means of Eqs. (14) and (15) of BTZ99. The evolution of the tachocline thickness 
has been considered, by assuming that the angular velocity at the base of the convective 
envelope evolves according to the Skumanich law (Skumanich 1972) and introducing the 
dependence of the differential rotational rate on the rotation velocity derived by Donahue, 
Saar \& Baliunas (1996). According to helioseismic measurements (Basu, 1997), we adopt 
$h=0.05 R_\odot$ for the current-age tachocline thickness. As in BTZ99, we assume a 
Brunt-V\"aiss\"al\"a frequency $N^2=25$ $\mu$Hz and the angular velocity at the base of 
the solar photosphere $\Omega=0.415$ $\mu$Hz. 

\begin{figure}
   \centering
   \includegraphics[width=\columnwidth]{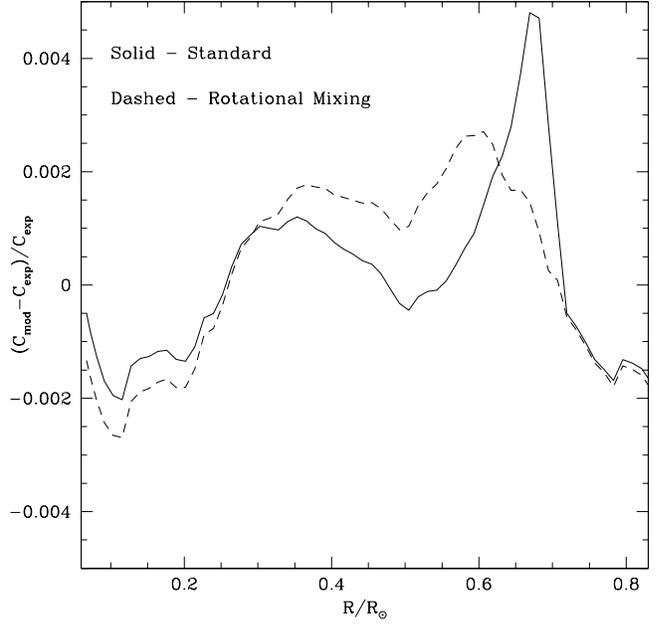}
   \caption{Sound-speed difference between GOLF+MDI data and theoretical models including 
            ({\it dashed line}) and not including ({\it solid line}) the rotational-induced 
            mixing in the tachocline (see text).  
           }
   \label{fig6}
\end{figure}

In Tab.~\ref{table5} we list some properties of the GN93-R model. The surface isotopic 
abundances and the $P$ parameters for $\element[][]{Li}$, $\element[][]{Be}$ and 
$\element[][]{B}$ are reported too. Owing to the introduction of this extra mixing, the 
diffusion efficiency of all the isotopes is significantly reduced (about $\sim 35\%$ less). 
As a consequence, a lower initial $Z$ is required in order to reproduce the present-day 
$Z/X$, thus leading to a larger $R_\mathrm{CE}$. Moreover, the predicted photospheric 
abundance of helium increases. As already discussed in BTZ99, the mixing in the tachocline 
smears off the sharp composition gradient just below the convective envelope, so that the 
resulting sound-speed profile becomes closer to the one derived from GOLF+MDI data (see 
Fig.~\ref{fig6}). However, in the intermediate zone ($0.3\le R/R_\odot\le 0.6$) the 
reproduction of the experimental sound-speed is less accurate. 

\begin{figure}
   \centering
   \includegraphics[width=\columnwidth]{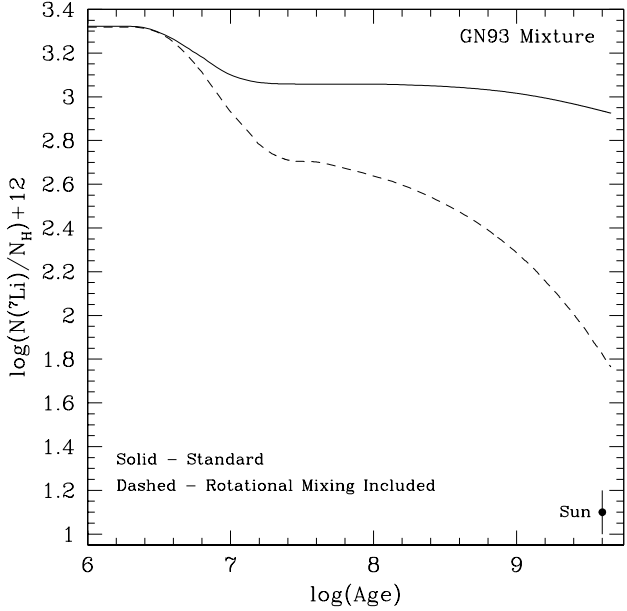}
   \caption{Evolution of the surface lithium abundance (astronomical scale) for models with 
            the same initial distribution of heavy elements (GN93 mixture), but computed 
            by including ({\it dashed lined}) or not including ({\it solid line}) the 
            rotational-induced mixing in the tachocline (see text). For comparison, the 
            solar photospheric value is also reported.
           }
   \label{fig7}
\end{figure}

The evolution of the photospheric abundance of $\element[][7]{Li}$ in GN93 and GN93-R models 
are compared in Fig.~\ref{fig7}. Note that the tachocline penetrates down to a layer where 
the temperature is high enough to burn $^7$Li but not enough to burn $\element[][9]{Be}$, 
$\element[][10]{B}$ and $\element[][11]{B}$. Nevertheless, the GN93-R SSM is not able to 
reproduce the observed lithium depletion. Indeed, although the resulting present-day 
photospheric abundance is substantially smaller than the one predicted by the GN93 model 
($\varepsilon(\mathrm{Li})=1.75$), it is still larger than the observed one. 

Similar conclusion was found by BTZ99 (model $B_t$), even if they obtained a larger lithium 
depletion that is marginally compatible with the observed value. In any case, 
the lithium depletion is largely increased when the rotational-induced mixing is included 
in the computation of SSMs. The exact value of the depletion efficiency depends on the 
adopted parameters for the mixing in the tachocline and the relevant physical inputs. 

\section{The absolute composition of stars}\label{sec5}

Once the solar composition is known, we can easily derive the absolute abundances in stars 
by using their spectroscopic determinations. 
Since these measurements are usually available for few elements only, in most cases just 
$\mathrm{[Fe/H]}$, stellar modelers commonly assume that all the other elements scale as iron. 
This is not the case of halo stars, for which an enhancement of the $\alpha$ elements 
($\element[][]{O}$, $\element[][]{Ne}$, $\element[][]{Mg}$, $\element[][]{Si}$, 
$\element[][]{S}$, $\element[][]{Ca}$, $\element[][]{Ti}$) with respect to iron is expected. 
In the following we will present the procedure to be used in these 
two common cases.

\subsection{Scaled solar compositions}

If, for a given star, we know $\mathrm{\left[Fe/H\right]}$ and if its composition is 
scaled-solar, we have:
\begin{equation}
\left(\frac{x_j}{x_\mathrm{Fe}}\right)_*=\left(\frac{x_j}{x_\mathrm{Fe}}\right)_\odot
\label{eq:11}
\end{equation}
\noindent
where the subscript $*$ refers to the star. By summing over the index $j$, it follows
\begin{equation}
\left(\frac{Z}{x_\mathrm{Fe}}\right)_*=\left(\frac{Z}{x_\mathrm{Fe}}\right)_\odot
\label{eq:12}
\end{equation}
\noindent
Being, by definition:
\begin{equation}
\mathrm{\left[\frac{Fe}{H}\right]}=\log\left(\frac{x_\mathrm{Fe}}{x_\mathrm{H}}\right)_*-\log\left(\frac{x_\mathrm{Fe}}{x_\mathrm{H}}\right)_\odot ,
\label{eq:13}
\end{equation}
\noindent
by combining Eq.~(\ref{eq:12}) and Eq.~(\ref{eq:13}), we find:
\begin{equation}
\mathrm{\left[\frac{Fe}{H}\right]}=\log\left(\frac{Z}{x_\mathrm{H}}\right)_*-\log\left(\frac{Z}{x_\mathrm{H}}\right)_\odot
\label{eq:14}
\end{equation}
\noindent
The last formula can be inverted to obtain the value of $Z_*$ corresponding to the measured 
$\mathrm{\left[Fe/H\right]}$:
\begin{equation}
Z_*=10^\mathrm{[Fe/H]} X_* \left(\frac{Z}{X}\right)_\odot
\label{eq:15}
\end{equation}
\noindent
if, in addition to $\mathrm{[Fe/H]}$, $X_*$ is known, or:
\begin{equation}
Z_*=\frac{10^\mathrm{[Fe/H]} \left(\frac{Z}{X}\right)_\odot (1-Y_*)}{1+10^\mathrm{[Fe/H]} \left(\frac{Z}{X}\right)_\odot}
\label{eq:16}
\end{equation}
\noindent
if $Y_*$ is known\footnote{Here $X=x_\mathrm{H}$ and $Y=x_\mathrm{He}$.}. Thus, the mass 
fraction of any element can be calculated by means of Eq.~(\ref{eq:10}).

Note that both the total metallicity ($Z$) and the element mass fractions ($x_j$) depend on 
$(Z/X)_\odot$. For example, if a star have $\mathrm{[Fe/H]}=-1$, its metallicity would be 
$Z=2.02\dot 10^{-3}$ or $Z=1.34\dot 10^{-3}$, depending on what solar composition is 
adopted, AG89 or Lo03\footnote{In both cases we have used $Y=0.245$.}, respectively.

\subsection{$\alpha$-enhanced compositions}

The distribution of heavy elements in halo stars may differ from a scaled solar composition. 
The most important difference regards the enhancement, with respect to iron, of the 
$\alpha$-elements. According to Salaris, Chieffi \& Straniero (1993), when the overabundance 
is similar for all the $\alpha$-elements, the total metallicity becomes:
\begin{equation}
Z_*=Z_0(a\cdot f_\alpha+b)
\label{eq:17}
\end{equation}
\noindent
where $Z_0$ is the metallicity calculated as in the case of a scaled solar composition (from 
Eq.~(\ref{eq:15}) or Eq.~(\ref{eq:16})), $f_\alpha=10^\mathrm{[\alpha/Fe]}$, 
$a=\sum_k(x_k/Z)_\odot$ ($k=\element[][]{O}$, $\element[][]{Ne}$, $\element[][]{Mg}$, 
$\element[][]{Si}$, $\element[][]{S}$, $\element[][]{Ca}$, $\element[][]{Ti}$) and $b=1-a$. 
In Tab.~\ref{table6} we report the coefficients $a$ and $b$ for the four different abundance 
compilations.
\begin{figure*}
   \centering
   \includegraphics[width=17.2cm]{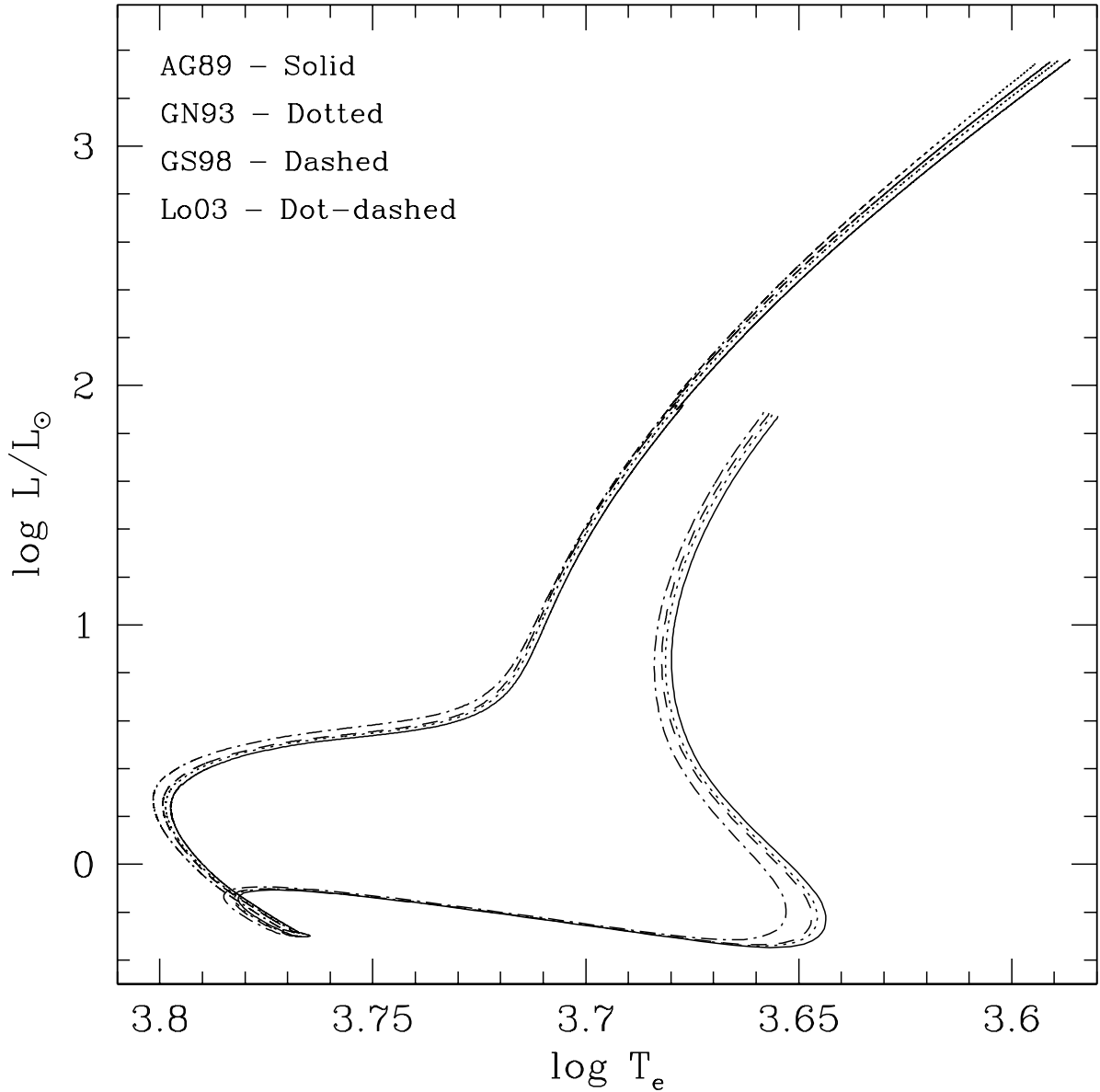}
   \caption{HR diagram for models with $M=0.8 M_\odot$, 
            $\left[\frac{\mathrm{Fe}}{\mathrm{H}}\right]=-1.4$ and
            $\left[\frac{\alpha}{\mathrm{Fe}}\right]=0.3$. For each adopted distribution 
            of heavy elements the values of $\alpha$ and $\mathrm{(Z/X)}_\odot$ have been 
            taken according to corresponding SSM (see text).
           }
   \label{fig8}
\end{figure*}

Also in this case, the metallicity of a given star depends on the adopted solar composition. 
The variation of $Z$ is mainly due to the variation of $(Z/X)_\odot$, while the changes of 
$a$ and $b$ have negligible consequences. This occurrence has many implications on the 
current interpretation of the observational data. As an example, Fig.~\ref{fig8} shows the 
evolutionary tracks for $M=0.8 M_\odot$, $\mathrm{[Fe/H]}=-1.4$, $\mathrm{[\alpha/Fe]}=0.3$ 
and $Y=0.245$, as obtained by assuming different solar compositions. Note that, although the 
total metallicity changes by a factor of 1.5, the effective temperature of the Main Sequence 
and of the RGB are only marginally affected. In this case, indeed, the variation of $Z$ is 
counterbalanced by the corresponding variation of the mixing length parameter. On the other 
hand, the luminosity of the Turn Off and that of the Subgiant Branch are sensitive to the 
variation of $Z$ only. When this effect is transported onto the isochrones, it affects the 
evaluation of the ages of Globular Clusters. By means of an age-metallicity relation (see 
Imbriani et al. 2004 for an update version), we find that the ages obtained with AG89 are 
$\sim 700$ Myr smaller than those obtained with Lo03. 

\begin{table}
\caption{Values of $a$ and $b$ parameters in Eq.~(\ref{eq:17}).}
\label{table6}
\centering
\begin{tabular}{lcc}
\hline\hline
Mixture & a & b \\
\hline
AG89    & 0.70057 & 0.29943 \\
GN93    & 0.68910 & 0.31090 \\
GS98    & 0.68728 & 0.31272 \\
Lo03    & 0.66946 & 0.33054 \\
\hline
\end{tabular}
\end{table}

\section{Summary and final remarks}

In this paper we have discussed the procedure to derive the absolute abundances of the
solar photosphere, of the early solar system (4.57 Gyr ago) and, in general, of stars, 
starting from the available data of relative elemental abundances. First of all, we have 
checked the validity of the fundamental assumption made to derive a complete set of elemental 
abundances starting from different incomplete data sets, as obtained from different solar 
system components. It is: the evolution of the Sun has occurred preserving the relative 
abundances of metals. We find that, although the absolute photospheric abundances of heavy 
elements decrease with time mainly due to microscopic diffusion, their relative ratios remain 
almost unaltered. Few exceptions concern lithium and elements having long-lived radioactive 
isotopes. However, these elements represent a very small fraction of the total metallicity, 
so that we can safely conclude that a unique value of the $R$ parameter can be used to 
combine photospheric and meteoritic data sets (see Eq.~(\ref{eq:3})). The value estimated 
by means of our SSMs is in very good agreement with the one obtained by comparing 
spectroscopic and meteoritic measurements of a selected sample of refractory elements.
The inclusion of radiative acceleration and a detailed evaluation of the partial ionization 
in the computation of the diffusion coefficients are not expected to modify the general 
conclusions of the present work, even if a more precise derivation of the absolute solar 
composition should deserve a proper treatment of these phenomena. 

Then, the absolute abundances of the solar photosphere and those of the early solar system 
have been obtained. In particular, the computed photospheric abundances can be used to 
evaluate the total metallicity and the absolute composition of stars for which the 
$\mathrm{\left[Fe/H\right]}$ and eventually $\left[\alpha/\mathrm{Fe}\right]$ have been 
measured.

Note that the procedure described in section 5 allows the determination of the current age 
photospheric values of the total metallicity and of the heavy elements absolute abundances 
of stars. These quantities have not to be confused with those of the material from which the 
star formed. The latter can be evaluated only doing additional assumptions about the 
modification of the photospheric chemical pattern, which differs from star to star, 
depending on the evolutionary status, the mass and the chemical composition.

Different compilations of solar abundances imply a different $Z-\mathrm{[Fe/H]}$ relation for 
stars having scaled solar composition. When, as usual, the input composition of a 
(scaled-solar) stellar model is compiled starting from the measured $\mathrm{[Fe/H]}$, it 
should be recalled that it depends on $(Z/X)_\odot$. Thus, the choice of a different solar
abundance compilation leads to different $Z$, even if the solar iron is not changed. 
Such an effect is particularly severe when passing from the classical AG89 compilation to 
the more recent Lo03. 

Recently Drake \& Testa (2005) have suggested that the solar abundance of Neon has to be 
increased. Such a conclusion is based on the evidence that nearby solar-like stars have an 
almost constant $\element[][]{Ne}/\element[][]{O}$ ratio which results about 2.7 times 
larger than the solar value, as determined from spectra of solar wind and solar corona 
(see also Cunha, Hubeny \& Lanz 2006). However, on the base of the analysis of solar active 
region spectra, Schmeltz et al. (2005) argued that such a ``enhanced-neon hypothesis'' is 
not founded for the Sun. In any case, if the photospheric solar neon abundance is increased 
according to Drake \& Testa, the corresponding $(Z/X)_\odot$ should increase up to 0.02016. 
By using this ``Ne-enhanced'' mixture in the computation of SSM, we obtain 
$Y_\mathrm{in}=0.27963$, $Z_\mathrm{in}=0.01653$, $Y_{\odot}=0.2477$, $Z_{\odot}=0.01487$ 
and $R_\mathrm{CE}=0.7238 R_\odot$. 
Let us finally mention that while we were working at the revision of the original manuscript, 
Antia \& Basu (2006) have presented an independent determination of the solar photospheric 
metallicity. By using the dimensionless gradient of the sound-speed, as derived by the 
inversion of solar oscillation frequencies, they obtain $Z_\odot=0.0172\pm 0.0002$, which 
is consistent with the theoretical value obtained by adopting the GS98 distribution of 
heavy elements. Unfortunately, the adopted method is only sensible to the total value of 
metallicity and it is largely independent on the relative elemental abundances, so that we 
can not draw definitive conclusion concerning the distribution of individual elements.
On the base of our models we can only confirm that SSMs computed by adopting the 
``old'' compilation of heavy elements abundances (from AG89 to GS98) better reproduce 
the physical properties of the current Sun as obtained by the helioseismic measurements.

\begin{acknowledgements}
      Special thanks to R. Gallino for the many enlightening discussions. We also thank the
      anonymous referee whose suggestions allowed us to greatly improve the paper.
      This work has been supported by the PRIN MIUR 2004 ``Nucleosynthesis in low mass 
      stars''. 
\end{acknowledgements}


\begin{thebibliography}{}

\bibitem[1994]{Alexander}
         Alexander, D.R., \& Ferguson, J.W., 1994, \apj, 437, 879
\bibitem[2001]{AllendePrieto01}
         Allende Prieto, C., Lambert, D.L., \& Asplund, M., 2001, \apjl, 556, L63
\bibitem[2002]{AllendePrieto02}
         Allende Prieto, C., Lambert, D.L., \& Asplund, M., 2002, \apjl, 573, L137
\bibitem[1989]{AndersGrevesse}
         Anders, E., \& Grevesse, N., 1989, \gca, 53, 197
\bibitem[1999]{Angulo}
         Angulo, C., and 27 co-authors, 1999, \nphysa, 656, 3
\bibitem[2006]{Antia}
         Antia, H.M., \& Basu, S., 2006, \apj, 644, 1292
\bibitem[2000]{Asplund}
         Asplund, M., 2000, \aap, 359, 755
\bibitem[2004]{AsplundetAl00}
         Asplund, M., Nordlund, A., Trampedach, R., \& Stein, R.F., 2000, \aap, 359, 743
\bibitem[2004]{AsplundetAl04}
         Asplund, M., Grevesse, N., Sauval, A.J., Allende Prieto, C., \& Kiselman, D.,
         2004, \aap, 417, 751
\bibitem[1998]{BahcalletAl98}
         Bahcall, J.N., Basu, S., \& Pinsonneault, M., 1998, Phys. Lett. B, 433, 128
\bibitem[2004]{BahcalletAl04}
         Bahcall, J.N., Serenelli, A.M., \& Pinsonneault, M., 2004, \apj, 614, 464
\bibitem[2005]{BahcalletAl05}
         Bahcall, J.N., Basu, S., Pinsonneault, M., \& Serenelli, A.M., 2005, \apj, 618, 1049
\bibitem[1997]{Basu97}
         Basu, S., 1997, \mnras, 288, 572
\bibitem[1997]{BasuAntia97}
         Basu, S., \& Antia, H.M., 1997, \mnras, 287, 189
\bibitem[2004]{BasuAntia04}
         Basu, S., \& Antia, H.M., 2004, \apjl, 606, L85
\bibitem[1997]{Bono97}
         Bono, G., Caputo, F., Cassisi, S., Castellani, V., \& Marconi, M., 1997, 
         \apj, 479, 279
\bibitem[2000]{Bono00}
         Bono, G., Caputo, F., Cassisi, S., Marconi, M., Piersanti, L., 
         \& Tornamb\'e, A., 2000, \apj, 543, 955
\bibitem[1999]{BrunetAl}
         Brun, A.S., Turck-Chi\`eze, S., \& Zahn, J.P., 1999, \apj, 535, 1032
\bibitem[1973]{Cameron73}
         Cameron, A.G.W., 1973, \ssr, 15, 121
\bibitem[1982]{Cameron82}
         Cameron, A.G.W., 1982, in Essays in Nuclear Astrophysics, ed. C.A. Barnes,
         D.D. Clayton \& D.N. Schramm (Cambridge Univ. Press), 23
\bibitem[1994]{Carlsson}
         Carlsson, M., Rutten, R.J., Brules, J.H.M.J., \& Shchukina, N.G., 1994,
         \aap, 288, 860
\bibitem[2002]{Casella}
         Casella, C., and 35 co-authors, 2002, \nphysa, 706, 203
\bibitem[1988]{Caughlan}
         Caughlan, G.R., \& Fowler, W.A., 1988, Atom. Data and Nucl. Data Tables, 40, 283
\bibitem[1998]{CharbonneletAl}
         Charbonnel, C., D\"appen, W., Schaerer, D., Bernasconi, P.A., Maeder, A., 
         Meynet, G., \& Mowlavi, N., 1999, \aaps, 135, 405
\bibitem[1998]{ChieffiLimongiStraniero}
         Chieffi, A., Limongi, M., \& Straniero, O., 1998, \apj, 502, 737
\bibitem[1889]{Clarke}
         Clarke, F.W., 1889, Bull. Phil. Soc. Washington, 11, 131
\bibitem[2000]{coc}
         Coc, A., Porquet, M.G., \& Nowacki, F., 2000, \prc, 61, 015801
\bibitem[1968]{CoxGiuli}
         Cox, J.P., \& Giuli, R.T., 1968, in Principles of Stellar Evolution, Vol. 1
         (New York: Gordon \& Breach), 281
\bibitem[2006]{CunhaetAl}
         Cunha, K., Hubeny, I., \& Lanz, T., 2006, astro-ph/0606738
\bibitem[1984]{DantonaMazzitelli}
         D'Antona, F., \& Mazzitelli, I., 1984, \aap, 138, 431
\bibitem[1973]{DewittetAl}
         Dewitt, H.E., Graboske, H.C., \& Cooper, M.S., 1973, \apj, 181, 439
\bibitem[1999]{Dominguez}
         Dominguez, I., Chieffi, A., Limongi, M., \& Straniero, O., 1999, \apj, 524, 226
\bibitem[1996]{DonahueetAl}
         Donahue, R.A., Saar, S.H., \& Baliunas, S.L., 1996, \apj, 466, 384
\bibitem[2005]{DrakeTesta}
         Drake, J.J., \& Testa, P., 2005, \nat, 436, 525
\bibitem[1995]{EstebanPeimbert}
        Esteban, C., \& Peimbert, M., 1995, \rmxaa \ SC, 3, 133
\bibitem[2004]{Luna}
         Formicola, A., and 31 co-authors, 2004, Phys. Lett. B, 591, 61
\bibitem[1998]{Girardi}
         Girardi, L., \& Bertelli, G., 1998, \mnras, 300, 533, 
\bibitem[1991]{GodeetAl}
        Goode, P.R., Dziembowski, W.A., Korzennik, S.G., \& Rhodes, E.J. Jr, 1991, 
        \apj, 367, 649
\bibitem[1973]{GraboskeetAl}
         Graboske, H.C., Dewitt, H.E., Grossman, A.S., \& Cooper, M.S., 1973, \apj, 181, 457
\bibitem[1984]{Grevesse}
         Grevesse, N., 1984, Phys. Scr., 8, 49
\bibitem[1993]{GrevesseNoels}
         Grevesse, N., \& Noels, A., 1993, in Origin and evolution of the elements,
         ed. N. Prantzos, E. Vangioni-Flam, \& M. Cass\'e, Cambridge
         University Press, 15
\bibitem[1998]{GrevesseSauval}
         Grevesse, N., \& Sauval, A.J., 1998, \ssr, 85, 161
\bibitem[1996]{Iglesias}
         Iglesias, C.A., \& Rogers, F.J., 1996, \apj, 464, 943
\bibitem[2004]{ImbrianietAl}
         Imbriani, G., and 29 co-authors, 2004, \aap, 420, 625
\bibitem[1977]{ItohetAl77}
         Itoh, N., Totsuji, H., \& Ichimaru, S., 1977, \apj, 218, 477
\bibitem[1979]{ItohetAl79}
         Itoh, N., Totsuji, H., Ichimaru, S., \& Dewitt, H.E., 1979, \apj, 234, 1079
\bibitem[2004]{Kim}
         Kim, Y.C., Demarque, P., Yi, S.K., \& Alexander, D.R., 2002, \apjs, 
         143, 499
\bibitem[1997]{KosovichevetAl.}
         Kosovichev, A.G., and 33 co-authors, 1997, \solphys, 170, 43
\bibitem[1997]{LazreketAl}
         Lazrek, M., and 19 co-authors, 1997, \solphys, 175, 227
\bibitem[2003]{Lodders}
         Lodders, K., 2003, \apj, 591, 1220
\bibitem[2001]{Maeder}
         Maeder, A., \& Meynet, G., 2001, \aap, 373, 555 
\bibitem[1990]{MihalasetAl}
         Mihalas, D., Hummer, D.G., Mihalas, B.W., \& Daeppen, W., 1990, \apj, 350, 300
\bibitem[1994]{oda}
         Oda, T., Hino, M., Muto, K., Takahara, M., \& Sato, K.,
         1994, Atom. Data Nucl. Data Tables, 56, 231
\bibitem[1993]{PalmeBeer}
         Palme, H., \& Beer, H., 1993, in Landolt B\"ornstein Group VI,
         Astronomy and Astrophysics, Vol. 2A, Ed. H.H. Voigt, Springer (Berlin), 196
\bibitem[2004]{Palme}
         Palme, H., \& Jones, A., 2004, in  Meteorites, Comets and Planets (Treatise on
         Geochemistry, vol. 1), ed. A.M. Davis, Elsevier (Amsterdam), 4
\bibitem[2004]{Pietrinferni}
         Pietrinferni, A., Cassisi, S., Salaris, M., \& Castelli, F., 2004, \apj, 
         612, 168 
\bibitem[1989]{Pinsonneaultetal}
         Pinsonneault, M.H., Kawaler, S.D., Sofia, S., \& Demarque, P., 1989, \apj, 338, 424
\bibitem[1998]{PolsetAl}
         Pols, O.R., Schroder, K.P., Hurley, J.R., Tout, C.A., \& Eggleton, P.P., 
         1998, \mnras, 298, 525
\bibitem[2002]{Prada}
         Prada Moroni, P.G., \& Straniero, O., 2002, \apj, 581, 585
\bibitem[1996]{RichardetAl}
         Richard, O., Vauclair, S., Charbonnel, C., \& Dziembowski, W.A., 1996, 
         \aap, 312, 1000
\bibitem[1992]{Rogers}
         Rogers, F.J., \& Iglesias, C.A., 1992, \apjs, 79, 507
\bibitem[1996]{Rogers2}
         Rogers, F.J., Swenson, F.J., \& Iglesias, C.A., 1996, \apj, 456, 902
\bibitem[1993]{SalarisetAl}
         Salaris, M., Chieffi, A., \& Straniero, O., 1993, \apj, 414, 580
\bibitem[1997]{Salari97}
         Salaris, M., degl'Innocenti, S., \& Weiss, A., 1997, \apj, 479, 665
\bibitem[2005]{SchmelzetAl.}
         Schmelz, J.T., Nasraoui, K., Roames, J.K., Lippner, L.A., \& Garst, J.W., 
         2005, \apjl, 634, L197
\bibitem[1972]{Skumanich}
         Skumanich, A., 1972, \apj, 171, 565
\bibitem[2003]{Spergel}
         Spergel, D.N., and 16 co-authors, 2003, \apjs, 148, 175
\bibitem[1992]{SpiegelZahn}
         Spiegel, E.A., \& Zahn, J.P., 1992, \aap, 265, 106
\bibitem[1988]{Straniero88}
         Straniero, O., 1988, \aaps, 76, 157
\bibitem[1997]{Straniero97}
         Straniero, O., Chieffi, A., \& Limongi, M., 1997, \apj, 490, 425
\bibitem[1956]{SuessUrey}
         Suess, H.E., \& Urey, H.C., 1956, Rev.Mod.Phys., 28, 53
\bibitem[1994]{thoul}
         Thoul, A.A., Bahcall, J.N., \& Loeb, A., 1994, \apj, 421, 828
\bibitem[1998]{Turcotte}
         Turcotte, S., Richer, J., Michaud, G., Iglesias, C.A. \& Rogers, F.J., 1998, 
         \apj, 504, 539
\end{thebibliography}
\end{document}